\documentclass[letterpaper,english,reprint, aps]{revtex4}
\usepackage[T1]{fontenc}
\usepackage{color}
\usepackage{verbatim}
\usepackage{float}
\usepackage{mathrsfs}
\usepackage{amsmath}
\usepackage{amssymb}
\usepackage{graphicx}
\usepackage{wasysym}

\makeatletter

\pdfpageheight\paperheight
\pdfpagewidth\paperwidth

\providecommand{\tabularnewline}{\\}

\@ifundefined{textcolor}{}
{%
 \definecolor{BLACK}{gray}{0}
 \definecolor{WHITE}{gray}{1}
 \definecolor{RED}{rgb}{1,0,0}
 \definecolor{GREEN}{rgb}{0,1,0}
 \definecolor{BLUE}{rgb}{0,0,1}
 \definecolor{CYAN}{cmyk}{1,0,0,0}
 \definecolor{MAGENTA}{cmyk}{0,1,0,0}
 \definecolor{YELLOW}{cmyk}{0,0,1,0}
}

\makeatother

\usepackage{babel}
\begin{document}
\preprint{APS/123-QED}
\title{RBG-Maxwell Framework: Simulation of Collisional Plasma Systems via
Coupled Boltzmann-Maxwell equations on GPU }

\author{Ming-Yan Sun}
\affiliation{Xi'an Research Institute of High Tech, Xi'an 710021, China,}
\author{ Peng Xu }
\affiliation{Xi'an Research Institute of High Tech, Xi'an 710021, China.}
\author{Jun-Jie Zhang$^{*}$ }
\affiliation{Division of Computational Physics and Intelligent Modeling, Northwest Institute of Nuclear Technology, Xi'an 710024, China,}
\author{Qun Wang}
\affiliation{Department of Modern Physics, University of Science and Technology of China, Hefei 230026, China}
\author{Tai-Jiao Du}
\affiliation{Division of Computational Physics and Intelligent Modeling, Northwest Institute of Nuclear Technology, Xi'an 710024, China.}
\author{Jian-Guo Wang$^{*}$}
\affiliation{School of Information and Communications Engineering, Xi'an Jiaotong University, Xi'an 710049, China}
\date{\today}
\thanks{corresponding authors: Jian-Guo Wang E-mail address: wanguiuc@mail.xjtu.edu.cn; \\
Jun-Jie Zhang E-mail address: zjacob@mail.ustc.edu.cn; \\
To use our code, please refer to https://Juenjie.github.io or https://sunminmgyan.github.io }

\begin{abstract}
This paper presents the RBG-Maxwell framework, a relativistic collisional
plasma simulator on GPUs. We provide detailed discussions on the fundamental
equations, numerical algorithms, implementation specifics, and key
testing outcomes. The RBG-Maxwell framework is a robust numerical
code designed for simulating the evolution of plasma systems through
a kinetic approach on large-scale GPUs. It offers easy adaptability
to a wide range of physical systems. Given the appropriate initial
distributions, particle masses, charges, differential cross-sections,
and external forces (which are not confined to electromagnetic forces),
the RBG-Maxwell framework can direct the evolution of
a particle system from a non-equilibrium state to a thermal state.
\end{abstract}
\maketitle

\section{Introduction}

As the forth state of matter along with the solid, liquid and gas,the
plasma comprises 99\% of the visible universe, ranging from the quark-gluon
matter at microscopic scales to plasma at macroscopic scales \citep{Verboncoeur2005}.
The self-consistent interaction of charged particles with electromagnetic
(EM) fields is essential to describe many plasma systems such as the
early universe \citep{Fukushima2016}, plasma \citep{Sang-Yun2022,Singh2022}, Tokamak \citep{Lijiangang2013,Lijiangang2021},
high-altitude nuclear explosion \citep{James2013,Wang2023,Brandon2021}, vacuum electronic devices \citep{Wang2010,Wang2018,Wang2013}, system
generated EM pulses \citep{DANIEL1978,JiannanChen2020,Chen_2022}, and
solar plasma, etc.


The self-consistent plasma model involves the classical EM fields
governed by Maxwell equations and the particle distributions in
coordinate space (phase space) governed by conservation (kinetic or
Boltzmann) equations. These equations are coupled to each other, namely,
the particles are sources to the fields while the fields exert forces
on the particles. In some cases, the particles can have radiations
(including $\alpha$, $\beta$ or $\gamma$ rays depending on energy
scales) in quantum transition processes. These particles and radiations
as well as the EM fields all interact with one another both classically
and in quantum processes, forming a complex system of particles and
fields at a wide range of energy scales. We take a plasma of ionized
hydrogen and electrons as an example. The mass of an $\text{H}^{+}$
ion is about 1836 times that of an electron, but they have the same
electric charge. Thus, the two species are at extremely different
space-time (or energy) scales under the influence of the same EM fields:
electrons can be easily accelerated even to the speed of light while
it is much harder for ions. This multi-scale feature is one of the
major challenges in a theoretical description of plamsa system \citep{Treumann2009}.


Due to the multi-scale feature and many physical processes in some
plasma systems, most theorectical models are focused on specific phemomena
at some particlular scales. In general these models can be put into
three categories: fluid dynamics models coupled with Maxwell equations
\citep{Shen2016,Kissmann2018,Ziegler1999}, the kinetic ion models
with ions treated as particles and electrons as fluids \citep{HEWETT2011,Peng2021},
and the fully kinetic models \citep{Daniel2018,Daniel2019,Xu2015,Xu2017}.
The general-purpose plasma toolkits also emerged in recent years beyond
the fields in which they had originally been developed \citep{Kushner2009,Van2009}.


There are mainly two numerical methods to solve the coupled Boltzmann
equations and Maxwell equations self-consistently: the particle
simulation method (e.g., the Particle-In-Cell method \citep{Robert2022,Maximilian2022,Arber2015,Jianguo2009,Dawson1983,CHEN20223415}
and the test particle method \citep{Grishmanovskii2022,Norman1964,Thomas1973,Xu2015},
etc.) and the direct numerical method \citep{Yan2021,Junjie2022,Dongxin2018,Richard1995}.
The two approaches have their own pros and cons in some circumstances.
For instance, the particle simulation method has an advantage in describing
complex moving geometries and high-deformation flows, while the direct
numerical method is more suitable for large scale and collisional
plasma systems. Since the latter is numerically more challenging than
the former, not many studies have been carried out in this direction.


The full numerical solution to the Boltzmann equation (BE) has always
been a computational challenge due to its high-dimension collisional
integrals even on today\textquoteright s petascale CPU clusters \citep{Dimarco2014,Romatschke2011}.
With the advent of Graphic Processing Unit (GPU) technology in parallel
computing, some calculations become feasible \citep{Werner2020,Januszewski2014,Pelusi2022,Xiaoyu2022,Jaiswal2019}.
In this paper, we focus on the direct numerical method to solve the
Relativistic Boltzmann equations on GPUs coupled with Maxwell equations
(the RBG-Maxwell framework) for collisional plasma.


The RBG-Maxwell framework contains following main modules as building
blocks. 

\CIRCLE{} A module for calculating the collision terms of the BE.
The evaluation of collision terms is made by ZMCintegral, a high-dimension
integration package based on GPU \citep{Wu:2019tsf,Zhang2019}. All
interactions among particles and radiations are incorporated into
collision terms through the matrix elements for particle scatterings
in quantum theory \citep{Peskin2018,PeskinMichael2018}.


\CIRCLE{} A module for calculating the drift terms of the BE. The
drift term in the left hand side of the BE describes the variation
of the particle's distribution in phase space due to its velocity
and the force exerted on it. The latter is also called the Vlasov
term which describes the drift of the particle from classical fields
such as the EM and gravitational fields.


\CIRCLE{} A module for calculating the EM fields. We solve Jefimenko's
equations for the EM fields as functions of space and time. The source
terms for Jefimenko's equations are determined from the distribution
functions of charged particles, whose evolution is governed by the
BE. An open source version of this module, JefiGPU \citep{Junjie2022b},
can be found and executed on the Code Ocean platform.


\CIRCLE{} A module to couple the BE to Maxwell equations. The distribution
functions of charged particles at one time play as inputs (the source
terms) to JefiGPU to give the EM fields at the same time which play
as inputs back to the BE to obtain the distribution functions at the
next time step. The loop continues till the end of the evolution at
the final time.


The purpose of the RBG-Maxwell framework is to build a general simulation
toolkit for collisional plasma. It can be applied to various plasma
systems such as pre-equilibrium state of quark-gluon plasma in high
energy heavy-ion collisions \citep{Kurkela2019a,Keegan2016,Kurkela2019,Kurkela2019b,Zhang2020},
solar plasma in collisions \citep{Imada2021,Todor2021}, collisional
plasma involving quantum states \citep{Jieru2020,Bu-Bo2021,yongtaozhao2021}
or particle annihilation/production/ionization \citep{Pasechnik2017},
etc..


\section{Unit and conventions}

\subsection{Natural unit}

Throughout this work, we use the natural unit (NU) with the reduced
Planck constant $\hbar$, the speed of light $c$, the vacuum permitivity
$\epsilon_{0}$ and the Boltzmann constant $k_{B}$ as basis units.
In different scenarios, we may choose different numerical values for
$\hbar$, $c$, $k_{B}$ and $\epsilon_{0}$ that are appropriate
for specific phenomena of interest at different energy scales. The
dimension of a physical quantity can be expressed as $\hbar^{\alpha}c^{\beta}k_{B}^{\gamma}\epsilon_{0}^{\delta}$.
For example, the proton's charge $e$ (electron's charge without sign),
which is $1.06\times10^{-19}$ Coulomb in the international unit system
(SI), is $e=0.302862\sqrt{\hbar c\epsilon_{0}}$ in NU, being obtained
from the fine structure constant $\alpha=e^{2}/(4\pi\epsilon_{0}\hbar c)\approx1/137$.
Any physical quantity in NU has the dimension of the energy to some
power if we set $\hbar=c=k_{B}=\epsilon_{0}=1$. This is called the
NU convention. Normally, the energy unit is MeV in NU or NU convention,
for example, the electron mass is 0.511 MeV and one second is $1.5\times10^{21}$
$\text{MeV}^{-1}$, etc..


In order to show how to convert a quantity between NU and SI, we start
with the values of $\hbar$ and $c$ in SI,
\begin{eqnarray}
\hbar & = & 1.05457\times10^{-34}\text{J\ensuremath{\cdot}s},\nonumber \\
c & = & 2.99792\times10^{8}\text{m/s}.\label{eq:hbarc}
\end{eqnarray}
In NU, we define an energy unit E to corresponding to Joules in SI
with a dimensionless ratio $\lambda=$E/J or $1\ \text{E}=\lambda\ \mathrm{J}$.
Note that E can span from MeV for microscopic processes to Joules
for macroscopic ones. Then we obtain 
\begin{equation}
\hbar c=3.16152\times10^{-26}\text{J\ensuremath{\cdot}m}=\frac{1}{\lambda}3.16152\times10^{-26}\text{\text{E}\ensuremath{\cdot}m}.\label{eq:conversion}
\end{equation}
From the above relation we convert the unit of length in SI (meter)
to NU 
\begin{eqnarray}
1\text{m} & = & 3.16303\times10^{25}\hbar c\lambda\ \text{E}^{-1}.\label{eq:unit_conversion}
\end{eqnarray}
Table \ref{tab:Conversion-tabel-between} shows the conversion between
SI and NU.


\begin{table}
\caption{Conversion between SI and NU. \label{tab:Conversion-tabel-between}}

\centering{}%
\begin{tabular}{|c|c|c|}
\hline 
quantity & SI & NU\tabularnewline
\hline 
length & m & $3.16304\times10^{25}\hbar c\lambda$ $\text{E}^{-1}$\tabularnewline
\hline 
mass & kg & $8.89752\times10^{16}/(\lambda c^{2})$ E\tabularnewline
\hline 
time & s & $9.48253\times10^{33}\hbar\lambda$ $\text{E}^{-1}$\tabularnewline
\hline 
momentum & kg$\cdot$m/s & $2.99792\times10^{8}/(\lambda c)$ E\tabularnewline
\hline 
Energy & J & $(1/\lambda)\ \text{E}$\tabularnewline
\hline 
force & kg$\cdot$m/$\text{s}^{2}$ & $3.16153\times10^{-26}/(\lambda^{2}\hbar c)$ $\text{E}^{2}$\tabularnewline
\hline 
temperature & K & $1.38065\times10^{-23}/(\lambda k_{B})$ E\tabularnewline
\hline 
electric charge & C & $1.89032\times10^{18}\sqrt{\hbar c\epsilon_{0}}$\tabularnewline
\hline 
proton's charge & 1.60218$\times10^{-19}$C & $0.30286\sqrt{\hbar c\epsilon_{0}}$\tabularnewline
\hline 
electric current & A & $1.99347\times10^{-16}\sqrt{c\epsilon_{0}}/(\lambda\sqrt{\hbar})$
E\tabularnewline
\hline 
electric field & V/m & $1.67249\times10^{-44}/(\lambda^{2}\hbar^{3/2}c^{3/2}\epsilon_{0}^{1/2})$
$\text{E}^{2}$\tabularnewline
\hline 
magnetic field & T & $5.01398\times10^{-36}/(\lambda^{2}\hbar^{3/2}c^{5/2}\epsilon_{0}^{1/2})$
$\text{E}^{2}$\tabularnewline
\hline 
\end{tabular}
\end{table}

\subsection{Conventions}

RBG-Maxwell is a toolkit for relativistic plasma and can also be
applied to non-relativistic plasma systems. The Latin letters in boldface
denote the spatial and momentum three-vectors in Cartesian coordinates,
e.g. $\mathbf{x}=(x,y,z)$ and $\mathbf{p}=(p_{x},p_{y},p_{z})$.
We choose the Greek letters such as $\mu$ to denote the time and
space time index with $\mu=0,1,2,3$. For example, the four-momentum
of particle with the rest mass $m$ is denoted as $p^{\mu}=(p^{0},\mathbf{p})=(E_{p}/c,\mathbf{p})$
or $p_{\mu}=(p_{0},-\mathbf{p})=(E_{p}/c,-\mathbf{p})$, where $E_{p}=c\sqrt{\mathbf{\left|p\right|}^{2}+m^{2}c^{2}}$
is the energy of the particle. The convention of the metric tensor
is $g^{\mu\nu}=\textrm{diag}(1,-1,-1,-1)$. The mass-shell condition
is $m^{2}c^{2}=g^{\mu\nu}p_{\mu}p_{\nu}=p^{\nu}p_{\nu}$, where the
Einstein's convention for summation is implied: repeated indices mean
summation, $p^{\nu}p_{\nu}\equiv\sum_{\nu=0,1,2,3}p^{\nu}p_{\nu}$.
Similar to the four-momentum $p^{\mu}$, the four-coordinate is denoted
as $x^{\mu}=(x^{0},\mathbf{x})=(ct,\mathbf{x})$. The Latin letter
$a$ denotes the particle species, such as electrons, protons, ions,
etc.. We also use the Latin letter $i=1,2,3$ to denote three spatial
components, for example, $x^{i}$ refers to $\mathbf{x}=(x^{1},x^{2},x^{3})=(x,y,z)$.


\section{RBG-Maxwell framework}

\subsection{Single-particle distribution function}

In relativistic kinetic theory, the single-particle distribution function
$f_{a}(t,\mathbf{x},\mathbf{p})$ for the particle species $a$ in
phase space is defined as 
\begin{equation}
f_{a}(t,\mathbf{x},\mathbf{p})=\frac{\Delta N_{a}}{\Delta V_{\text{phase}}}=\frac{\Delta N_{a}}{\Delta x^{3}\Delta p^{3}},\label{eq:discrete_f}
\end{equation}
where $\Delta N_{a}$ is the number of particles in the phase-space
volume $\Delta V_{\text{phase}}=\Delta x^{3}\Delta p^{3}$. In Eq.
(\ref{eq:discrete_f}), all particles in the volume $\Delta V_{\text{phase}}$
are labeled by the same $\mathbf{x}$ and $\mathbf{p}$. From $f_{a}(t,\mathbf{x},\mathbf{p})$,
we can obtain the particle number density $n_{a}(t,\mathbf{x})$ and
the particle number $N_{a}(t)$ for the particle species $a$, 
\begin{eqnarray}
n_{a}(t,\mathbf{x}) & = & \int d^{3}\mathbf{p}f_{a}(t,\mathbf{x},\mathbf{p}),\nonumber \\
N_{a}(t) & = & \int d^{3}\mathbf{x}n_{a}(t,\mathbf{x}),\label{eq:normalization}
\end{eqnarray}
as well as the particle current density 
\begin{eqnarray}
j_{a}(t,\mathbf{x}) & = & \int d^{3}\mathbf{p}\mathbf{v}f_{a}(t,\mathbf{x},\mathbf{p}),\label{eq:particle flow}
\end{eqnarray}
with $\mathbf{v}=\mathbf{p}/p_{0}=c\mathbf{p}/E_{p}$ being the relativistic
velocity for the particle species $a$. We can obtain the total quantities
by summation over the particle species $a$
\begin{equation}
n(t,\mathbf{x})=\sum_{a}n_{a}(t,\mathbf{x}),\;N(t)=\sum_{a}N_{a}(t),\;j(t,\mathbf{x})=\sum_{a}j_{a}(t,\mathbf{x}).
\end{equation}
The particle number conservation equation reads 
\begin{equation}
\frac{\partial n(t,\mathbf{x})}{\partial t}+\nabla\cdot j(t,\mathbf{x})=0.
\end{equation}
The total energy-momentum tensor can be obtained via
\begin{eqnarray}
T^{\mu\nu}(x) & = & \sum_{a}c\int\frac{d^{3}\mathbf{p}}{p^{0}}p^{\mu}p^{\nu}f_{a}(t,\mathbf{x},\mathbf{p}),\label{eq:energy momentum tensor}
\end{eqnarray}
where $T^{00}(x)$ is the energy density, $T^{0k}(x)=T^{k0}(x)$ with
$k=1,2,3$, is the momentum density (energy flux across the surface
perpendicular to the $k$-direction), and $T^{ij}$ with $i,j=1,2,3$,
is the momentum flux in the $i$-th component across the surface perpendicular
to the $j$-direction. In particular, $T^{ii}(x)$ is called the normal
stress, and $T^{ij}(x)$ with $i\neq j$ is called the shear stress.
Then the energy-momentum conservation without EM fields can be expressed
as 
\begin{eqnarray}
\partial_{\mu}T^{\mu\nu}(x) & = & 0,\label{eq:current conservation}
\end{eqnarray}
where a summation over $\mu$ is implied. Other quantities, such as
the electrical conductivity, the shear and bulk viscosity, can be
obtained from single-particle distribution functions similarly. 


In many systems such as hydro-magneto fluids, we often use the single-particle
distribution in local thermal equilibrium such as the Maxwell-Boltzmann
distribution 
\begin{equation}
f_{a}^{\text{LTE}}(t,\mathbf{x},E_{p})\sim\exp\left[-\frac{E_{p}-\mu_{a}}{k_{B}T(t,\mathbf{x})}\right],
\end{equation}
where $\mu$ is the chemical potential for the particle species $a$,
and $T(t,\mathbf{x})$ is the local temperature. The off-equilibrium
distribution functions can be obtained by solving the BE as shown later. 


\subsection{Coupled Boltzmann-Maxwell equations }

The BE describes the evolution of the particle's phase space distribution
in a particle system from an off-equilibrium state to the equilibrium
state. The BE is just the equation that the rate of the variation
in the phase space distribution function is caused by particle collisions.
For the particle species $a$ with the distribution $f_{a}(t,\mathbf{x},\mathbf{p})$,
the BE reads 
\begin{eqnarray}
\frac{df_{a}(t,\mathbf{x},\mathbf{p})}{dt} & = & C[f_{a}(t,\mathbf{x},\mathbf{p})],\label{eq:BE1}
\end{eqnarray}
where $C[f_{a}(t,\mathbf{x},\mathbf{p})]$ is the collision term and
$df_{a}/dt$ in the left-hand-side is given by 
\begin{eqnarray}
\frac{df_{a}(t,\mathbf{x},\mathbf{p})}{dt} & = & \frac{\partial f_{a}(t,\mathbf{x},\mathbf{p})}{\partial t}+\frac{\partial\mathbf{x}}{\partial t}\cdot\nabla_{\mathbf{x}}f_{a}(t,\mathbf{x},\mathbf{p})+\frac{\partial\mathbf{p}}{\partial t}\cdot\nabla_{\mathbf{p}}f_{a}(t,\mathbf{x},\mathbf{p})\nonumber \\
 & = & \frac{\partial f_{a}(t,\mathbf{x},\mathbf{p})}{\partial t}+\mathbf{v}\cdot\nabla_{\mathbf{x}}f_{a}(t,\mathbf{x},\mathbf{p})+\mathbf{F}\cdot\nabla_{\mathbf{p}}f_{a}(t,\mathbf{x},\mathbf{p}),\label{eq:total_diff}
\end{eqnarray}
where $\partial\mathbf{x}/\partial t=\mathbf{v}$ and $\partial\mathbf{p}/\partial t=\mathbf{F}$
are respectively the effective velocity and force of the particle.
$\nabla_{\mathbf{x}}$ and $\nabla_{\mathbf{p}}$ are the differential
operators in coordinate and momentum space respectively. One can obtain
$\partial\mathbf{x}/\partial t$ and $\partial\mathbf{p}/\partial t$
from the equation of motion of the particle. The terms proportional
to $\partial\mathbf{x}/\partial t$ and $\partial\mathbf{p}/\partial t$
in Eq. (\ref{eq:total_diff}) are usually called the drift and Vlasov
terms, respectively.


The collision term describes short-range particle scatterings which
can be calculated from the first principle such as quantum theory.
In a plasma systems, the force is the long-range EM force, $\mathbf{F}=Q_{a}(\mathbf{E}+\mathbf{v}_{a}\times\mathbf{B})$,
where $Q_{a}$ is the electric charge of the particle species $a$,
and $\mathbf{E}$ and $\mathbf{B}$ are the electric and magnetic
fields acting on the particle, respectively. If there are no collisions
among particles, i.e. the collision term is zero, the BE is called
the Vlasov equation. It describes the time evolution of the collisionless
system of charged particles in the long-range EM field. The EM field
can be solved from Maxwell equations, 
\begin{eqnarray}
\nabla\times\mathbf{B} & = & \frac{1}{c^{2}}\frac{\partial\mathbf{E}}{\partial t}+\frac{1}{c^{2}\epsilon_{0}}\mathbf{J}\nonumber \\
\nabla\times\mathbf{E} & = & -\frac{\partial\mathbf{B}}{\partial t}\nonumber \\
\nabla\cdot\mathbf{B} & = & 0\nonumber \\
\nabla\cdot\mathbf{E} & = & \frac{\rho}{\epsilon_{0}},\label{eq:Maxwell}
\end{eqnarray}
where $\epsilon_{0}$ and $c$ can take arbitrary values in NU. In
the above equations, $\rho$ is the electric charge density, $\mathbf{J}$
is the electric current density, and they are related to the particle's
distribution functions solved from the BE
\begin{eqnarray}
\rho & = & \sum_{a}\int d^{3}\mathbf{p}f_{a}(t,\mathbf{x},\mathbf{p})Q_{a},\nonumber \\
\mathbf{J} & = & \sum_{a}\int d^{3}\mathbf{p}f_{a}(t,\mathbf{x},\mathbf{p})Q_{a}\mathbf{v}_{a}.\label{eq:source_term}
\end{eqnarray}


\subsection{Collision term in BE\label{subsec:Collision-term-for}}

In this subsection, we discuss the construction of the collision term.
The collision term can be expressed as momentum integrals for the
binary collision (with the momentum setup) $a(\mathbf{k}_{1})+b(\mathbf{k}_{2})\rightarrow c(\mathbf{k}_{3})+d(\mathbf{p})$
for particles $a,b,c$, and $d$ 
\begin{eqnarray}
C_{ab\rightarrow cd} & \equiv & \int\prod_{i=1}^{3}d^{3}\mathbf{k}_{i}[f_{\mathbf{k}_{1}}^{a}f_{\mathbf{k}_{2}}^{b}F_{\mathbf{k}_{3}}^{c}F_{\mathbf{p}}^{d}-F_{\mathbf{k}_{1}}^{a}F_{\mathbf{k}_{2}}^{b}f_{\mathbf{k}_{3}}^{c}f_{\mathbf{p}}^{d}],\nonumber \\
 &  & \times\delta^{(4)}(k_{1}+k_{2}-k_{3}-p)\frac{\hbar^{2}c|M_{ab\leftrightarrow cd}|^{2}}{64\pi^{2}k_{1}^{0}k_{2}^{0}k_{3}^{0}p^{0}},\label{eq:C_a_b}
\end{eqnarray}
where $M_{ab\leftrightarrow cd}$ is the invariant matrix element
which can be obtained from microscopic theories such as quantum theory.
If microscopic degree of freedoms (DOF) are involved such as spin
and color in Quantum Chromodynamics \citep{Greiner2007,Weinberg1995},
an average over the DOF in the initial state and a sum over the DOF
in the final state are implied in $M_{ab\leftrightarrow cd}$, and
a sum over the DOF of the distribution function is also implied. The
delta function in Eq. (\ref{eq:C_a_b}) 
\begin{eqnarray}
\delta^{(4)}(k_{1}+k_{2}-k_{3}-p) & = & \delta^{(3)}(\mathbf{k}_{1}+\mathbf{k}_{2}-\mathbf{k}_{3}-\mathbf{p})\nonumber \\
 &  & \times\delta(k_{1}^{0}+k_{2}^{0}-k_{3}^{0}-p^{0}),\label{eq:delta}
\end{eqnarray}
ensures the energy and momentum conservation in the binary collision.
The quantum correction 
\begin{eqnarray}
F_{\mathbf{p}}^{d} & \equiv & 1+\theta\overline{n}(t,\mathbf{x},\mathbf{p}),\label{Quantum_correction}
\end{eqnarray}
where $\theta=\pm1,0$ for the Bose-Einstein, Fermi-Dirac, and Boltzmann
statistics, respectively. The dimensionless quantity $\overline{n}(t,\mathbf{x},\mathbf{p})=(2\pi\hbar)^{3}f(t,\mathbf{x},\mathbf{p})/g$
is called the occupation number with $g$ being the number of DOF.
For electrons, we take $g=2$ for the spin DOF; for photon or $X$-ray,
we take $g=1$; for quarks and gluons, we take $g=6$ and $g=16$, respectively. 


We note that the collision term (\ref{eq:C_a_b}) is only valid when
the momenta of colliding particles are independent of their positions
and are uncorrelated before collisions. In history, the assumption
is called by Boltzmann as the ``Stosszahlansatz'' (molecular chaos
hypothesis) \citep{Sandri1966,Ehrenfest1959}. Without such an assumption,
the correlation between colliding particles have to be considered
instead of using $f_{\mathbf{k}_{1}}^{a}f_{\mathbf{k}_{2}}^{b}$ in
Eq. (\ref{eq:C_a_b}). The Stosszahlansatz is deeply related to the
Boltzmann's H-theorem \citep{Plastino2001,Boltzmann2003}.


In a plasma system, there are some different collisions. For example,
a photon may collide with a hydrogen atom to make it ionized, i.e.
$\gamma+\text{H}\rightarrow e^{-}+\text{H}^{+}$. Its reverse reaction
is the combination of an electron and a hydrogen ion to make a hydrogen
atom with emission of a photon: $e^{-}+\text{H}^{+}\rightarrow\text{H}+\gamma$.
The Hydrogen atom may exchange an electron with another ion, i.e.,
$\text{H}+\text{H}^{+}\rightarrow\text{H}+\text{H}^{+}$. Meanwhile,
a hydrogen ion can collide with one another, i.e., $\text{H}^{+}+\text{H}^{+}\rightarrow\text{H}^{+}+\text{H}^{+}$.
So the collision term for $\text{H}^{+}$ is

\begin{equation}
\begin{array}{ccc}
C[f_{\text{H}^{+}}(t,\mathbf{x},\mathbf{p})] & = & \left[\frac{1}{2}C_{\text{H}^{+}+\text{H}^{+}\rightarrow\text{H}^{+}+\text{H}^{+}}\right.\\
 &  & +C_{\gamma+\text{H}\rightarrow\text{H}^{+}+e^{-}}\\
 &  & +\left.C_{\text{H}+\text{H}^{+}\rightarrow\text{H}+\text{H}^{+}}\right].
\end{array}\label{eq:collisionFe}
\end{equation}
Since all $\text{H}^{+}$ are indistinguishable, we have introduced
the symmetry factor 1/2 in front of the term $C_{\text{H}^{+}+\text{H}^{+}\rightarrow\text{H}^{+}+\text{H}^{+}}$
when all incident particles are of the same type (species).


\subsection{Cross section and matrix element}

Sometimes we only have the measured cross sections in experiments
instead of matrix elements. So it is necessary to find the relationship
between the cross section and the matrix element.

For a two-to-two scattering process $ab\leftrightarrow cd$, the differential
cross section $d\sigma$ is defined through the differential probability
per unit time 
\begin{eqnarray}
d\omega & = & c(2\pi\hbar)^{4}\delta^{(4)}\left(k_{1}+k_{2}-k_{3}-p\right)|M_{ab\leftrightarrow cd}|^{2}V\nonumber \\
 &  & \times\frac{1}{2E_{1}2E_{2}2E_{3}2E_{p}}\frac{\hbar^{4}c^{4}}{V^{4}}\frac{Vd^{3}\mathbf{k}_{3}}{(2\pi\hbar)^{3}}\frac{Vd^{3}\mathbf{p}}{(2\pi\hbar)^{3}}\nonumber \\
 & = & \frac{cI}{k_{1}^{0}k_{2}^{0}V}d\sigma,\label{eq:dods}
\end{eqnarray}
where $V$ is the space volume, $k_{1}=(E_{1},\mathbf{k}_{1})$ and
$k_{2}=(E_{2},\mathbf{k}_{2})$ are the energy-momenta of incoming
particles respectively, and $k_{1}=(E_{1},\mathbf{k}_{1})$ and $k_{2}=(E_{2},\mathbf{k}_{2})$
are the energy-momenta of outgoing particles respectively. The invariant
flux $I$ of incoming particles is defined as 
\begin{eqnarray}
I(k_{1}^{\mu},k_{2}^{\mu}) & = & \sqrt{(k_{1}^{\mu}k_{2,\mu})^{2}-m_{1}^{2}m_{2}^{2}c^{4}}.\label{eq:invariant_flux}
\end{eqnarray}
Combining Eqs. (\ref{eq:dods}) and (\ref{eq:invariant_flux}),
we obtain the differential cross section in the Lab frame for the
particle $d$, 
\begin{equation}
\frac{d\sigma}{d\Omega}\vert_{a(\mathbf{k}_{1})+b(\mathbf{k}_{2})\leftrightarrow c(\mathbf{k}_{3})+d(\mathbf{p})}=\frac{\hbar^{2}|M_{ab\leftrightarrow cd}|^{2}|\mathbf{p}|}{64\pi^{2}k_{3}^{0}I(k_{1}^{\mu},k_{2}^{\mu})}.\label{eq:ds_M_2_2}
\end{equation}


\subsection{Jefimenko's equations}

As a general solution to Maxwell equations, Jefimenko's equations
\citep{Jefimenko1989} can be directly derived from the retarded potential
in Maxwell equations \citep{Max1920,Shao2016}, 
\begin{eqnarray}
\mathbf{E}(\mathbf{r},t) & = & \frac{1}{4\pi\epsilon_{0}}\int\left[\frac{\mathbf{r}-\mathbf{r}^{\prime}}{|\mathbf{r}-\mathbf{r}^{\prime}|^{3}}\rho(\mathbf{r}^{\prime},t_{r})+\frac{\mathbf{r}-\mathbf{r}^{\prime}}{|\mathbf{r}-\mathbf{r}^{\prime}|^{2}}\text{\ensuremath{\frac{1}{c}}}\frac{\partial\rho(\mathbf{r}^{\prime},t_{r})}{\partial t}\right.\nonumber \\
 &  & \left.-\frac{1}{|\mathbf{r}-\mathbf{r}^{\prime}|}\text{\ensuremath{\frac{1}{c^{2}}}}\frac{\partial\mathbf{J}(\mathbf{r}^{\prime},t_{r})}{\partial t}\right]d^{3}\mathbf{r}^{\prime},\label{eq:E}\\
\mathbf{B}(\mathbf{r},t) & = & -\frac{1}{4\pi\epsilon_{0}c^{2}}\int\left[\frac{\mathbf{r}-\mathbf{r}^{\prime}}{|\mathbf{r}-\mathbf{r}^{\prime}|^{3}}\times\mathbf{J}(\mathbf{r}^{\prime},t_{r})\right.\nonumber \\
 &  & +\left.\frac{\mathbf{r}-\mathbf{r}^{\prime}}{|\mathbf{r}-\mathbf{r}^{\prime}|^{2}}\times\text{\ensuremath{\frac{1}{c}}}\frac{\partial\mathbf{J}(\mathbf{r}^{\prime},t_{r})}{\partial t}\right]d^{3}\mathbf{r}^{\prime},\label{eq:B}
\end{eqnarray}
where $t_{r}=t-|\mathbf{r}-\mathbf{r}^{\prime}|/c$ is the retarded
time, $\mathbf{E}$ and $\mathbf{B}$ are the electric and magnetic
fields at the space-time point $(\mathbf{r},t)$, respectively, and
$\rho$ and $\mathbf{J}$ are the charge and current densities at
the space-time point $(\mathbf{r}^{\prime},t_{r})$, respectively.


The integration form in Eqs. (\ref{eq:E}) and (\ref{eq:B}) has some merits.
First, different from the finite-difference time-domain (FDTD) method
\citep{Yee1966}, Jefimenko's equations do not rely on boundary conditions.
In the FDTD method, choosing proper boundary conditions is a complex
and technical sub-field \citep{William2004,ANGELL1992,ADLER2017,Brackbill2008,SHEEH1968}.
A slight change of boundary conditions can lead to different EM fields
\citep{Jean2012}. The integration form, on the other hand, merely
relys on the source terms, i.e. time-dependent charge and current
densities $\rho(\mathbf{r}^{\prime},t_{r})$ and $\mathbf{J}(\mathbf{r}^{\prime},t_{r})$.
Once one obtains $\rho(\mathbf{r}^{\prime},t_{r})$ and $\mathbf{J}(\mathbf{r}^{\prime},t_{r})$
from the BE through Eq. (\ref{eq:source_term}), it is much easier
to put the calculation of $\mathbf{E}(\mathbf{\mathbf{r}},t)$ and
$\mathbf{B}(\mathbf{r},t)$ through Jefimenko's equations (\ref{eq:E}) and (\ref{eq:B})
on GPU clusters than through Maxwell equations. However, there is
a cost to pay in using the integration form. In the FDTD method, $\mathbf{E}(\mathbf{\mathbf{r}},t)$
and $\mathbf{B}(\mathbf{r},t)$ in the vicinity $(\mathbf{\mathbf{r}}\pm d\mathbf{\mathbf{r}},t\pm dt)$
are needed, while in Eqs. (\ref{eq:E}) and (\ref{eq:B}) the sources $\rho(\mathbf{r}^{\prime},t_{r})$
and $\mathbf{J}(\mathbf{r}^{\prime},t_{r})$ at all previous time
in the entire computational domain are needed.


Moreover, the integrations in Jefimenko's equations are always well-defined
except there is a divergence at $\mathbf{r}=\mathbf{r}^{\prime}$.
The near-source divergence has its profound physical origins \citep{Peskin2018,PeskinMichael2018}.
The usual treatment of the singularity is to adopt the finite size
of the spatial grid as a cut-off, $|\mathbf{r}-\mathbf{r}^{\prime}|_{\text{cut-off}}=\sqrt{(dx)^{2}+(dy)^{2}+(dz)^{2}}$.
When $|\mathbf{r}-\mathbf{r}^{\prime}|<|\mathbf{r}-\mathbf{r}^{\prime}|_{\text{cut-off}}$
we can set $\rho(\mathbf{r}^{\prime},t_{r})=\mathbf{J}(\mathbf{r}^{\prime},t_{r})=0$,
so the divergence is removed. In this case, the EM field in a specific
grid can only be generated by the charge and current densities in
other grids.


\section{Algorithms }

\begin{figure}
\begin{centering}
\includegraphics[scale=0.3]{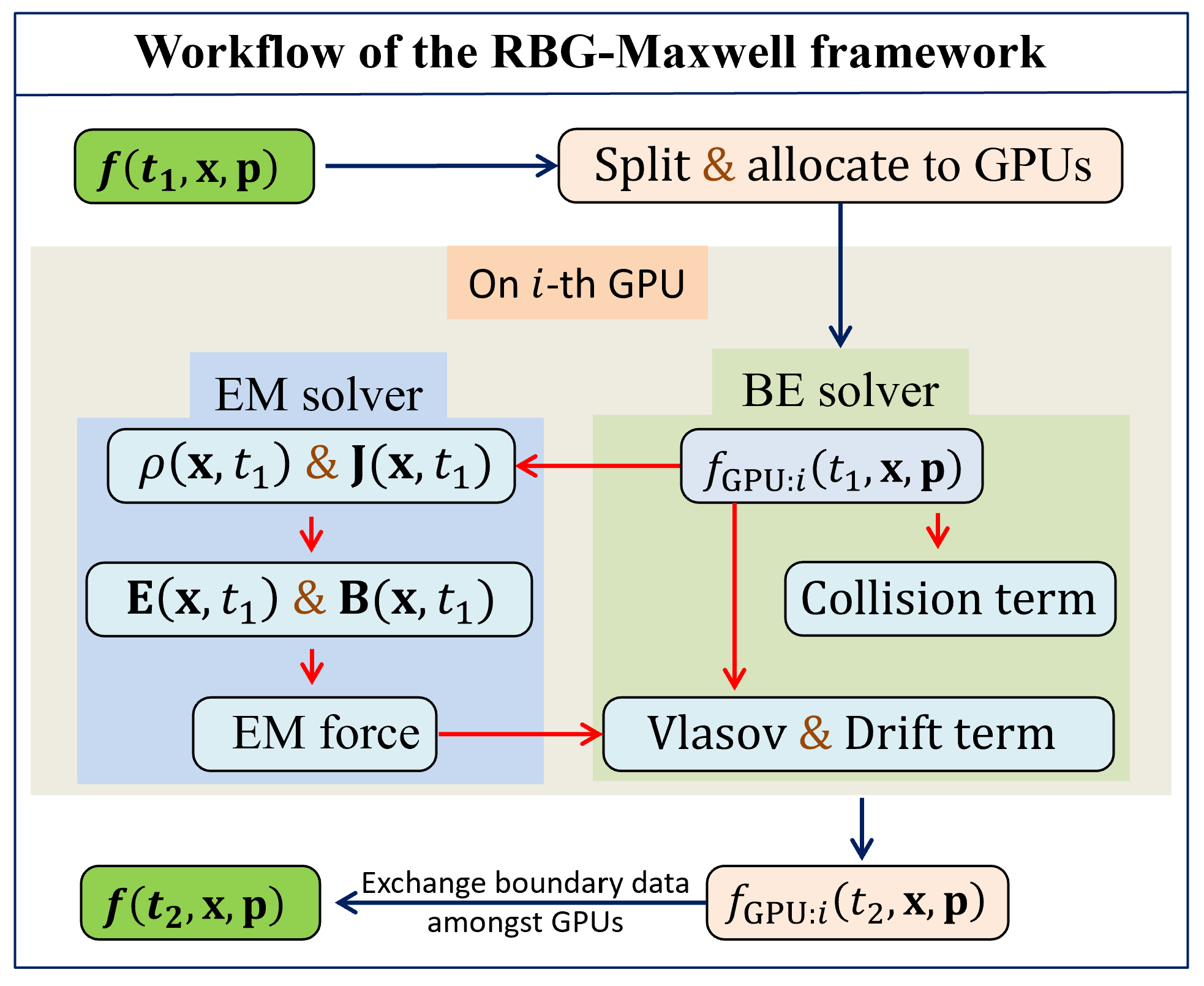}
\par\end{centering}
\caption{Schematic of the workflow of the RBG-Maxwell framework. BE (Boltzmann)
and EM (EM) solvers are revised from our previous packages JefiGPU
\citep{Junjie2022b} and RBG\citep{Zhang2020}. At each time step,
BE solver provides the charge and current densities to the EM solver,
and EM solver gives the EM forces excerted on the particles. \label{fig:Schematic-of-the-1}}
\end{figure}

The RBG-Maxwell framework is a first-principle based plasma toolkit.
In the current version, limited by the molecular chaos hypothesis
\citep{Sandri1966,Ehrenfest1959}, it can only deal with dilute plasma
systems at the weak coupling limit. Similar to most plasma toolkits,
it contains two main blocks in general, as shown in Fig. \ref{fig:Schematic-of-the-1},
the BE solver for the phase space distribution functions and the EM
field solver for the EM fields generated by these distributions.


As a stable solution to Maxwell equations, Jefimenko's equations
give the EM fields from charge and current densities provided by distribution
functions. Given the distribution $f(t_{1},\mathbf{x},\mathbf{p})$
and the external force $\mathbf{F}(t_{1},\mathbf{x},\mathbf{p})$
depending on the EM field at a previous time $t_{1}$, the BE solver
gives $f(t_{2},\mathbf{x},\mathbf{p})$ at a later time $t_{2}$.
In the BE solver, the collision term $\mathscr{\mathfrak{\mathcal{C}}}[f]$
is performed in the Monte Carlo approach with the symmetric sampling
method \citep{Zhang2020}. The drift and Vlasov terms $\mathbf{v}_{a}\cdot\nabla_{\mathbf{x}}$
and $\mathbf{F}\cdot\nabla_{\mathbf{p}}$ are differentiated by the
first or second order finite difference method. Then $f(t_{2},\mathbf{x},\mathbf{p})$
is input to Jefimenko's equations to obtain the EM fields at $t_{2}$
from which a new cycle is started. Therefore the RBG-Maxwell framework
is a fully consistent algorithm for the coupled relativistic Boltzmann-Maxwell
equations.


All modules are implemented by Python and are made heavily parallel
on GPU clusters, which makes a fast solver to the coupled equations.
All quantities in RBG-Maxwell are consistently defined on GPU clusters
via the Python package CuPy \citep{cupy_learningsys2017,Cupy} (CuPy
is an implementation of NumPy-compatible multi-dimensional array on
CUDA). The CUDA kernel functions are written with the Python package
Numba \citep{Lam2015,Numba} (an open source JIT compiler that translates
a subset of Python and NumPy code into fast machine code).

\vspace{1.5em}

\subsection{Boltzmann Equation solver}

The BE solver is an updated version of our previous work \textendash{}
Relativistic Boltzmann equations on GPUs (RBG) \citep{Zhang2020}.
The algorithm for the calulation of the collision term is the same
as in RBG, while other parts are redesigned to fit the RBG-Maxwell
framewrok. In the calculation of distribution functions, we must always
keep them non-negative, otherwise, the algorithm will quickly blow
up and lead to wrong solutions. In order to preserve their positivity,
we need to revise some of existing positivity preversing schemes accordingly.
First and second order positivity preversing upwind difference schemes
are adopted to approximate spatial gradients, and first order positivity
preserving upwind difference scheme is used to approximate momentum
gradients. Other functionalities, such as non-negative distribution
functions and flux limiters, are also included in the framework.


For numerical convenience, we rewrite Eqs. (\ref{eq:BE1}) and (\ref{eq:total_diff})
into three equations
\begin{eqnarray}
\frac{1}{3}\frac{\partial f_{a}(t,\mathbf{x},\mathbf{p})}{\partial t}+\mathbf{v}\cdot\nabla_{\mathbf{x}}f_{a}(t,\mathbf{x},\mathbf{p}) & = & 0,\label{eq:BE_part1}\\
\frac{1}{3}\frac{\partial f_{a}(t,\mathbf{x},\mathbf{p})}{\partial t}+\mathbf{F}\cdot\nabla_{\mathbf{p}}f_{a}(t,\mathbf{x},\mathbf{p}) & = & 0,\label{eq:BE_part2}\\
\frac{1}{3}\frac{\partial f_{a}(t,\mathbf{x},\mathbf{p})}{\partial t} & = & C[f_{a}(t,\mathbf{x},\mathbf{p})].\label{eq:BE_part3}
\end{eqnarray}
Therefore, solving the BE is equivalent to solving above three equations.


\subsubsection{Drift term}

We use the first and second order finite difference methods in calculating
distribution functions. For a fast and smooth evolution of distribution
functions, one is advised to use the first order upwind difference
scheme, but with the problem of significant diffusion \citep{Nishikawa2010,Farhad1995,Davis1976}
which is problematic for relativistic plasma systems since it breaks
causality to a large extent. To overcome such a problem, we have to
employ the second order flux-limited method to suppress the diffusion.


\paragraph*{First order scheme.}

We adopt the unconditionally positivity preserving finite difference
scheme \citep{Benito2013,Matthias2014} (UPFD) to approximate the
gradients of drift terms. The usual treatment of gradients is 
\begin{eqnarray}
v_{x}\frac{\partial f_{a}(t,x,y,z)}{\partial x} & = & v_{x}\frac{f_{a}(t,x,y,z)-f_{a}(t,x-\Delta x,y,z)}{\Delta x}\ \ (v_{x}>0),\nonumber \\
v_{x}\frac{\partial f_{a}(t,x,y,z)}{\partial x} & = & v_{x}\frac{f_{a}(t,x+\Delta x,y,z)-f_{a}(t,x,y,z)}{\Delta x}\ \ (v_{x}<0).\label{eq:first_order_upwind}
\end{eqnarray}
The upwind difference does not gaurantee that $f_{a}(t+\Delta t,\mathbf{x})$
is always positive when updated via Eq. (\ref{eq:BE_part1}). To achieve
positivity, we can make following replacements, 
\begin{eqnarray}
f_{a}(t,x,y,z)-f_{a}(t,x-\Delta x,y,z) & \rightarrow & f_{a}(t+\Delta t,x,y,z)-f_{a}(t,x-\Delta x,y,z)\ \ (v_{x}>0),\nonumber \\
f_{a}(t,x+\Delta x,y,z)-f_{a}(t,x,y,z) & \rightarrow & f_{a}(t,x+\Delta x,y,z)-f_{a}(t+\Delta t,x,y,z)\ \ (v_{x}<0).\label{eq:UPFD}
\end{eqnarray}
Substituting Eq. (\ref{eq:UPFD}) into Eq.(\ref{eq:BE_part1}) we
obtain an explicit expression for always positive distributions
\begin{eqnarray}
f_{a}(t+\Delta t,\mathbf{x}) & = & \text{\ensuremath{\frac{3\Delta x}{3|v_{x}|\Delta t+\Delta x}}}\left(\frac{1}{3}f_{a}(t,\mathbf{x})+\frac{\Delta t|v_{x}|}{\Delta x}\begin{cases}
f_{a}(t,x-\Delta x,y,z) & v_{x}>0\\
f_{a}(t,x+\Delta x,y,z) & v_{x}<0
\end{cases}\right).\label{eq:explicit_UPDF_drift}
\end{eqnarray}
The same formula can be obtained in the $y$ and $z$ directions.


\paragraph*{Second order scheme.}

As will be illustrated in Sec. \ref{sec:Verification-on-multi-GPUs},
the first order upwind difference scheme brings significant numerical
diffusion. So we need to adopt the second order scheme to suppress
the numerical diffusion. The method we use in the RBG-Maxwell framework
is adapted from the positivity-preserving flux-limited method (PFL)
\citep{Bernard2011}.


For convenience, we use $f_{i}^{n}$ to denote $f_{a}(t,\mathbf{x},\mathbf{p})$,
where $n$ and $i=x,y,z$ represent time and spatial indices respectively.
Following the PFL, Eq. (\ref{eq:BE_part1}) can be approximated by
\begin{eqnarray}
\frac{1}{3}\frac{f_{i}^{n+1}-f_{i}^{n}}{\Delta t}+\sum_{i=x,y,z}\frac{F_{i+1/2}-F_{i-1/2}}{\Delta x_{i}} & = & 0,\label{eq:PFL}
\end{eqnarray}
where the convection flux is defined as 
\begin{eqnarray}
F_{i+1/2} & = & v_{i}^{+}f_{i}^{n}+\frac{1}{2}\phi_{i+1/2}^{+}(v_{i}^{+}f_{i}^{n}-v_{i-1}^{+}f_{i-1}^{n})\nonumber \\
 &  & +v_{i+1}^{-}f_{i+1}^{n}+\frac{1}{2}\phi_{i+1/2}^{-}(v_{i+1}^{-}f_{i+1}^{n}-v_{i+2}^{-}f_{i+2}^{n}).\label{eq:flux}
\end{eqnarray}
Here a flux limiter $\phi$ is introduced in front of the difference
terms. Its function is to decrease the value of the gradient and further
suppress the numerical diffusion. The flux limiters take the following
form
\begin{eqnarray}
\phi_{i+\frac{1}{2}}^{+} & = & \text{max}\left(0,\text{min}\left(1,\underbrace{\frac{v_{i+1}^{+}f_{i+1}^{n}-v_{i}^{+}f_{i}^{n}}{v_{i}^{+}f_{i}^{n}-v_{i-1}^{+}f_{i-1}^{n}}}_{\text{flux limiter for monotonicity preservation}},\underbrace{\frac{2}{\text{max}(\theta,(v_{i-1}^{+}f_{i-1}^{n}-v_{i}^{+}f_{i}^{n})/v_{i}^{+}f_{i}^{n})}}_{\text{additional limiter for positivity preservation}}\right)\right),\nonumber \\
\phi_{i+\frac{1}{2}}^{-} & = & \text{max}\left(0,\text{min}\left(1,\underbrace{\frac{v_{i}^{-}f_{i}^{n}-v_{i+1}^{-}f_{i+1}^{n}}{v_{i+1}^{-}f_{i+1}^{n}-v_{i+2}^{-}f_{i+2}^{n}}}_{\text{\text{flux limiter for monotonicity preservation}}},\underbrace{\frac{2}{\text{max}(\theta,(v_{i+2}^{-}f_{i+2}^{n}-v_{i+1}^{-}f_{i+1}^{n})/v_{i+1}^{-}f_{i+1}^{n})}}_{\text{\text{additional limiter for positivity preservation}}}\right)\right),\label{eq:flux_limiter}
\end{eqnarray}
where a Total Variation Diminishing (TVD) \citep{Anderson1986,Gottlieb1998}
limiter is used to preserve monotonicity and an additional limiter
is adopted to preserve positivity. We set the parameter $\theta=10^{-10}$
throughout the code.


\paragraph*{Boundary conditions.}

In real cases, the particles can be either absorbed or reflected by
the boundaries. For practical convenience, we introduce a positive
parameter $\zeta(x,y,z)$ {[}$0<\zeta(x,y,z)<1${]} to describe the
reflective probability and then $1-\zeta(x,y,z)$ represents the absorbtion
probability at the boundaries.


\subsubsection{Vlasov term}

\paragraph{First order scheme.}

Similar to the drift term, we also apply the UPDF scheme to the Vlasov
term,
\begin{eqnarray}
f_{a}(t+\Delta t,\mathbf{p}) & = & \text{\ensuremath{\frac{3\Delta p_{x}}{3|F_{x}|\Delta t+\Delta p_{x}}}}\left(\frac{1}{3}f_{a}(t,\mathbf{p})+\frac{\Delta t|F_{x}|}{\Delta p_{x}}\begin{cases}
f_{a}(t,p_{x}-\Delta p_{x},p_{y},p_{z}) & F_{x}>0\\
f_{a}(t,p_{x}+\Delta p_{x},p_{y},p_{z}) & F_{x}<0
\end{cases}\right).\label{eq:explicit_UPDF_Vlasov}
\end{eqnarray}
The same formula can be obtained in the $y$ and $z$ directions.

\paragraph{Boundary conditions.}

Given that the distribution functions invariably approach zero at
high momentum values, we can utilize periodic boundaries for differentiation
across all momentum dimensions. It's important to underscore that
this implementation will maintain the conservation of momentum intact,
despite any variations.

\subsubsection{Collision term}

The collision term presents the most significant challenge in the
numerical implementation of the Boltzmann Equation, as it necessitates
high-dimensional integration at each phase space grid. For binary
collisions, it is crucial to integrate the delta function prior to
any additional numerical implementations. To resolve the delta function,
we adopt the technique of integration over $d^{3}\mathbf{k}_{2}$
and $dk_{1z}$ as described by Ref. \citep{Zhang2020} in Eq. (\ref{eq:C_a_b}).
For a more detailed explanation, please refer to Appendix \ref{subsec:Introduction-of-the},
\begin{eqnarray}
 &  & \int\prod_{i=1}^{3}d^{3}\mathbf{k}_{i}\delta^{(4)}(k_{1}+k_{2}-k_{3}-p)\nonumber \\
 & = & \int d^{3}\mathbf{k}_{3}dk_{1}^{x}dk_{1}^{y}\sum_{i=\pm}\frac{1}{|J(k_{1z}^{i})|},\label{eq:work out delta E}
\end{eqnarray}
where the Jacobin 
\begin{eqnarray}
J(k_{1z}^{\pm}) & = & \frac{k_{1z}^{\pm}}{(k_{1}^{0})^{\pm}}-\frac{-k_{1z}^{\pm}+k_{3z}+p_{z}}{(k_{2}^{0})^{\pm}},\nonumber \\
k_{1z}^{\pm} & = & \mathrm{Root}[k_{1}^{0}+k_{2}^{0}-k_{3}^{0}-p^{0}=0].\label{eq:jacobin}
\end{eqnarray}
There are two roots for $k_{1z}$ from the equation $k_{1}^{0}+k_{2}^{0}-k_{3}^{0}-p^{0}=0$,
and $k_{1z}$ has the form of $k_{1z}^{\pm}\equiv\frac{C1\pm\sqrt{H}}{C2}$,
where $C1,C2,$ and $H$ are functions of $k_{1}^{x},k_{1}^{y}$ and $\mathbf{k}_{3}$.
The explicit expressions of $C1,C2,$ and $H$ are given in Appendix \ref{subsec:Finding-the-expression}.
Substituting Eq. (\ref{eq:work out delta E}) into Eq. (\ref{eq:C_a_b}),
we obtain the 5-dimensional collision integral
\begin{eqnarray}
C_{ab\rightarrow cd} & = & \int d^{3}\mathbf{k}_{3}dk_{1}^{x}dk_{1}^{y}\frac{\hbar^{2}c|M_{ab\leftrightarrow cd}|^{2}}{64\pi^{2}k_{1}^{0}k_{2}^{0}k_{3}^{0}p^{0}}\sum_{i=\pm}\frac{1}{|J(k_{1z}^{i})|}\nonumber \\
 &  & \times[f_{\mathbf{k}_{1}}^{a}f_{\mathbf{k}_{2}}^{b}F_{\mathbf{k}_{3}}^{c}F_{\mathbf{p}}^{d}-F_{\mathbf{k}_{1}}^{a}F_{\mathbf{k}_{2}}^{b}f_{\mathbf{k}_{3}}^{c}f_{\mathbf{p}}^{d}].\label{eq:5dcollision}
\end{eqnarray}
Eq. (\ref{eq:5dcollision}) can be evaluated numerically by the Direct
Monte Carlo (DMC) method\citep{LeBeau1999,DStefanov2019}. In RBG-Maxwell,
the 5-dimensional integration is performed by the GPU package ZMCintegral\citep{Zhang2019,Wu:2019tsf}.

\subsection{EM field solver}

A distinctive characteristic of a plasma system is the emergence of
(classical) Electromagnetic (EM) fields within the calculation domains.
These EM fields can originate from two potential sources: the fluctuating
or constant ambient fields generated by other systems, and the fields
produced by the motion of the plasma particles. A uniform GPU-based
approach to both the EM fields and the particle transport necessitates
a stable and concise solution to Maxwell equations. In our quest
for complete consistency in the relativistic limit, we have opted
for the integral form (specifically, Jefimenko's equation) of the
EM fields as opposed to the FDTD method.

For numerical convenience, we discretize Eqs. (\ref{eq:E}) and
(\ref{eq:B}) as
\begin{eqnarray}
\mathbf{E}(\mathbf{r},t) & = & \frac{1}{4\pi\epsilon_{0}}d\Omega^{\prime}\Sigma_{i,j,k}\left[\frac{\mathbf{r}-\mathbf{r}_{i,j,k}^{\prime}}{|\mathbf{r}-\mathbf{r}_{i,j,k}^{\prime}|^{3}}\rho(\mathbf{r}_{i,j,k}^{\prime},t_{r})\right.\nonumber \\
 &  & +\frac{\mathbf{r}-\mathbf{r}_{i,j,k}^{\prime}}{|\mathbf{r}-\mathbf{r}_{i,j,k}^{\prime}|^{2}}\text{\ensuremath{\frac{1}{c}}}\frac{\rho(\mathbf{r}_{i,j,k}^{\prime},t_{r})-\rho(\mathbf{r}_{i,j,k}^{\prime},t_{r}-dt)}{dt}\nonumber \\
 &  & \left.-\frac{1}{|\mathbf{r}-\mathbf{r}_{i,j,k}^{\prime}|}\text{\ensuremath{\frac{1}{c^{2}}}}\frac{\mathbf{J}(\mathbf{r}_{i,j,k}^{\prime},t_{r})-\mathbf{J}(\mathbf{r}_{i,j,k}^{\prime},t_{r}-dt)}{dt}\right]\nonumber \\
\label{eq:E-1}\\
\mathbf{B}(\mathbf{r},t) & = & -\frac{1}{4\pi\epsilon_{0}c^{2}}d\Omega^{\prime}\Sigma_{i,j,k}\left[\frac{\mathbf{r}-\mathbf{r}_{i,j,k}^{\prime}}{|\mathbf{r}-\mathbf{r}_{i,j,k}^{\prime}|^{3}}\times\mathbf{J}(\mathbf{r}_{i,j,k}^{\prime},t_{r})\right.\nonumber \\
 &  & +\left.\frac{\mathbf{r}-\mathbf{r}_{i,j,k}^{\prime}}{|\mathbf{r}-\mathbf{r}_{i,j,k}^{\prime}|^{2}}\times\text{\ensuremath{\frac{1}{c}}}\frac{\mathbf{J}(\mathbf{r}_{i,j,k}^{\prime},t_{r})-\mathbf{J}(\mathbf{r}_{i,j,k}^{\prime},t_{r}-dt)}{dt}\right]\nonumber \\
\label{eq:B-1}\\
t_{r} & = & t-\frac{|\mathbf{r}-\mathbf{r}_{i,j,k}^{\prime}|}{c},\label{eq:tr-1}
\end{eqnarray}
where index $i\in\{1,2,...,n_{x}\}$, $j\in\{1,2,...,n_{y}\}$, $k\in\{1,2,...,n_{z}\}$
and volume element $d\Omega^{\prime}=dx^{\prime}dy^{\prime}dz^{\prime}$.
$n_{x}$, $n_{y}$ and $n_{z}$ denote the number of spatial grids
used in the calculation. It's crucial to acknowledge that $\mathbf{r}$
and $\mathbf{r}_{i,j,k}^{\prime}$ can be defined in distinct regions.
The calculation domain of the EM solver comprises two areas: the source
region (which contains $\mathbf{r}_{i,j,k}^{\prime}$) and the observation
region (which houses $\mathbf{r}$ ). The source region is the domain
where the single particle distribution function is defined. The observation
region, on the other hand, may envelop the source region or encompass
other domains of interest. For instance, suppose a plasma clump is
produced in a laboratory, and we're interested in the electromagnetic
field it generates outside the lab. In this case, the source region
pertains to the cavity containing the plasma, and the observation
region includes both the source region and the area outside the lab.

The superiority of our method over the more efficient FDTD approach
manifests in two ways in RBG-Maxwell. Firstly, the integration method
is more stable compared to the FDTD approach. During numerical calculations,
integrations are less likely to yield infinities, which significantly
simplifies the GPU implementation. Secondly, the FDTD approach is
local, meaning the EM fields are determined only by the surrounding
fields. If the EM fields in the adjacent grids are not updated accurately,
the errors will accumulate over subsequent time steps. This is particularly
crucial in ultra-relativistic scenarios. On the contrary, the Jefimenko
equations are non-local - the EM fields are calculated considering
all possible grids across the entire spatial domain. Consequently,
they are less sensitive to the conditions of nearby grids compared
to the FDTD approach.

\subsection{Parallelization among clusters}

The essence of parallelization lies in minimizing the data exchange
volume among the GPU cards. The parallelization module in RBG-Maxwell
handles two components: the parallelization of single particle distribution
functions and that of electromagnetic fields. For distribution functions,
we partition the spatial domain into multiple sub-domains, which share
boundaries with each other. At each time step, only the shared boundaries
of the distribution functions are exchanged among the GPU cards. For
the electromagnetic fields $\mathbf{E}$ and $\mathbf{B}$, the distribution
functions on each GPU card yield the electric charge density $\rho$
and the electric current density $\mathbf{J}$. On each GPU card,
the EM fields produced by $\rho$ and $\mathbf{J}$ for the entire
spatial domain are calculated and distributed to the respective GPU
cards. Hence, only the computed EM fields are exchanged among the
GPU cards, while the source terms are retained on the GPU cards throughout
the simulation.

These functionalities are enabled by the Python package Ray\citep{PMoritz2018,Ray},
which provides a straightforward and universal API for developing
distributed applications.

\subsubsection{Ghost Cells in spatial domain}

The spatial domain is segmented into numerous sub-domains, dependent
on the total number of available GPU cards. To enhance clarity, we
utilize the symbol $\Omega$ to represent the domain of the spatial
region. In a GPU cluster comprising $M$ GPUs, $\Omega$ is partitioned
into $\Omega_{1},\Omega_{2},\cdots,\Omega_{M}$ sub-domains, each
corresponding to a separate GPU card.

\begin{figure}
\begin{centering}
\includegraphics[scale=0.45]{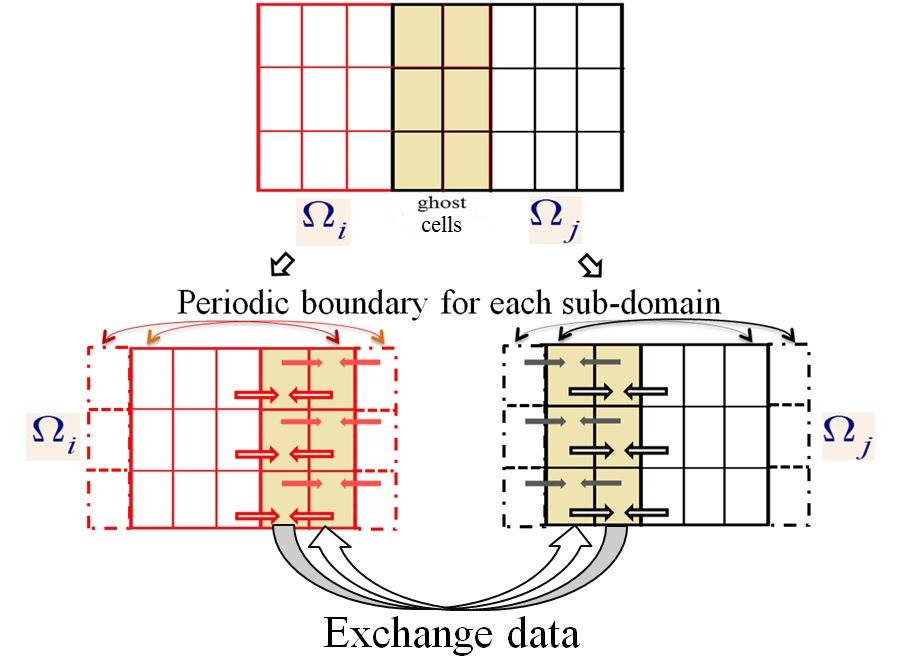}
\par\end{centering}
\caption{Illustration of the ghost cells between two overlapping regions. The
ghost cells between $\Omega_{i}$ (depicted by the red grid) and $\Omega_{j}$
(illustrated by the black grid) are shaded in yellow. In the first-order
upwind difference, the value of the distribution function at a spatial
grid is influenced by its nearest grids, as demonstrated by the mutually
pointing arrows. Given that we employ periodic boundaries for each
sub-domain, the distribution functions at the edges of the sub-domain
may not be accurate (as indicated by the regions pointed out by the
two mutually pointing solid arrows). \label{fig:Illustration-of-the}}
\end{figure}

The distribution function of a spatial grid located at the edges of
a sub-domain is influenced by its two adjacent grids - one of these
adjacent grids replicates from the opposite edge (regions connected
by the arc-arrows in Fig. \ref{fig:Illustration-of-the}). We apply
the concept of ghost cells (as explored by Refs. \citep{Lin1999,Keyes2000}),
a technique broadly employed for parallel computations. Ghost cells
are designated as the overlapped regions between two adjacent sub-domains,
as demonstrated in Fig. \ref{fig:Illustration-of-the}. After each
time step, the values of the distribution functions at the edges are
updated by exchanging data with the neighboring sub-domains.

\subsubsection{Parallelization of EM fields}

The division of the spatial domain for the electromagnetic (EM) fields
mirrors that of the distribution functions. On each GPU card (consequently
each sub-domain $\Omega_{i}$ where $i$ is in the set $\{1,2,\cdots,M\}$),
the integrations of the distribution functions yield the electric
current density$\mathbf{J}$ and the electric charge density $\rho$.
The calculated $\mathbf{J}$ and $\rho$ at each time snapshot will
be preserved in the GPU memory. Following the formulation of the Jefimenko's
equations (\ref{eq:E-1})\textasciitilde (\ref{eq:tr-1}), $\mathbf{J}$
and $\rho$ in region $\Omega_{i}$ contribute to the EM fields in
$\Omega_{1},\Omega_{2},\cdots,\Omega_{M}$.

\begin{figure}
\begin{centering}
\includegraphics[scale=0.5]{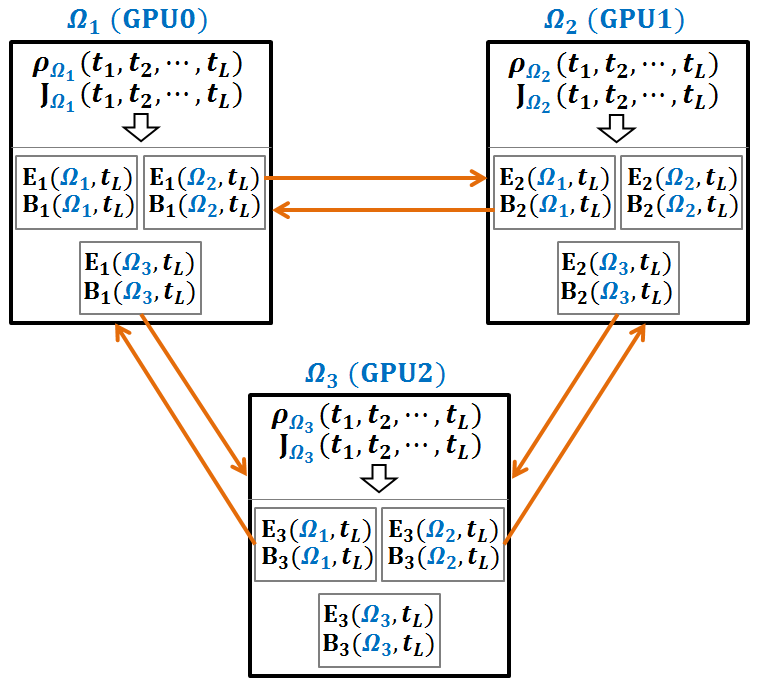}
\par\end{centering}
\caption{Parallelization scheme of the electromagnetic fields on GPUs. The
sources $\mathbf{J}$ and $\rho$ on each card give the EM fields
on all GPU cards. The obtained EM fields are distributed to other
GPU cards instantly. \label{fig:Illustration-of-the-1}}
\end{figure}

In Fig. \ref{fig:Illustration-of-the-1}, we illustrate the parallelization
scheme for a case comprising three sub-domains $\Omega=\{\Omega_{1},\Omega_{2},\Omega_{3}\}$.
The time sequence $\{t_{1},t_{2},\cdots,t_{L}\}$ represents the evaluated
time steps, with the current time step being denoted as $t_{L}$.
The $\mathbf{J}$ and $\rho$ at each time snapshot are preserved
in the GPU memory, forming the sequences $\rho_{\Omega_{i}}(t_{1},t_{2},\cdots,t_{L})$
and $\mathbf{J}_{\Omega_{i}}(t_{1},t_{2},\cdots,t_{L})$ where $\Omega_{i}$
is a member of $\{\Omega_{1},\Omega_{2},\Omega_{3}\}$. Utilizing
$\rho_{\Omega_{i}}$ and $\mathbf{J}_{\Omega_{i}}$, the EM fields
for sub-domains $\Omega_{1},\Omega_{2},\Omega_{3}$ can be calculated
via Eqs. (\ref{eq:E-1})\textasciitilde (\ref{eq:tr-1}). The EM
fields thus obtained on each GPU card are then transferred to the
respective sub-domains. Throughout the evaluation process, only the
EM fields at the time step $t_{L}$ are exchanged. The resolution
(or equivalently, the number of spatial grids) for each sub-domain
$\Omega_{i}$ is constrained by the resolution of the Boundary Element
(BE) solver. At the current stage, the number of grids for a BE solver
on a single GPU card is approximately of size $[n_{x},n_{y},n_{z},n_{p_{x}},n_{p_{y}},n_{p_{z}}]\simeq[10,10,10,10,10,10]$.
Therefore, the exchanged EM fields have a size of $[n_{x},n_{y},n_{z}]\simeq[10,10,10]$.
With a GPU card boasting a few million cores, the exchanged EM fields
can reach a size of $[n_{x},n_{y},n_{z}]\simeq[100,100,100]$, which
equates to a few megabytes. Consequently, the volume of the exchanged
EM fields is rather small.

\textcolor{magenta}{}

In some cases, the plasma is restricted in a system while the interested
EM fields is outside the plasma system. In these senarios, each GPU
card gives the EM fields for the interested regions following a similarly
scheme as in Fig. \ref{fig:Illustration-of-the-1}.

\vspace{1.5em}

\section{Verification on multi-GPUs\label{sec:Verification-on-multi-GPUs}}

In this section, we will validate the RBG-Maxwell framework from three
different perspectives. Firstly, we employ a simple model example
to illustrate the evolution of particles and their respective electromagnetic
fields. The resulting outcomes are then compared with the plasma toolkit,
JefiPIC. Secondly, we select a particle system composed of two species
and confine these particles within a box. Over time, the particles
engage in collisions and gradually transition into a thermal state.
Lastly, we execute the code on eight GPU cards to assess the parallel
performance. During all these tests, the absorbing boundary conditions
are applied to all spatial boundaries.

\subsubsection{Test of drift and Vlasov terms}

We use a two dimensional (2D) pure electron plasma system to test
the drift and Vlasov terms. The 2D domain is divided into many spatial
grids with each grid having a size of $dx=dy=dz=10^{-5}\text{m}$.

\paragraph{Difference between first and second order schemes.}

\begin{figure}

\begin{centering}
\includegraphics[scale=0.4]{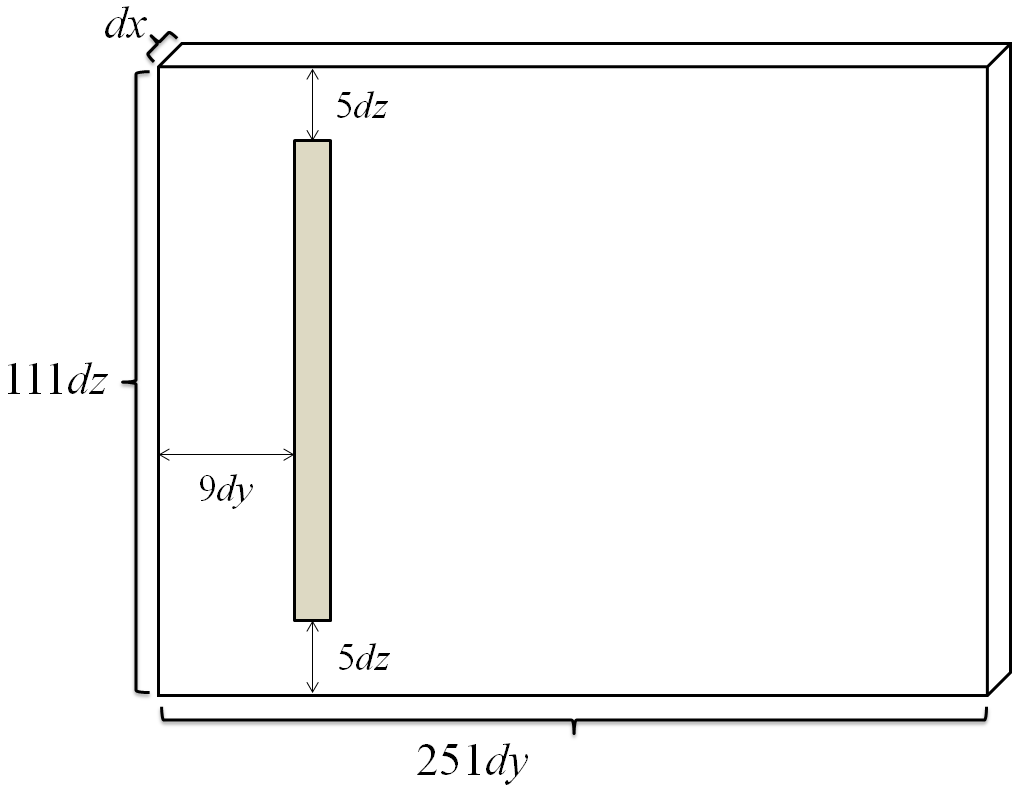}\caption{Schematic illustration of the initial configuration of the electron
plasma for order comparison. \label{fig:Schematic-illustration-of-1}}
\par\end{centering}
\end{figure}

As depicted in Fig. \ref{fig:Schematic-illustration-of-1}, the grid
numbers in the spatial domain are selected to be $n_{x}\times n_{y}\times n_{z}=1\times251\times111$.
A total of 31,250 electrons are uniformly dispersed across the shaded
regions, each with a constant initial velocity of $1.87683\times10^{6}\text{m}/\text{s}$.
This initial setup corresponds to the distribution function
\begin{eqnarray}
f_{e}(\mathbf{x}_{i},\mathbf{p}_{j},t_{0}) & = & \begin{cases}
\frac{31250}{101dV} & \mathbf{x}_{i}\ \text{in the shaded area \& }|\mathbf{p}_{j}|\leftrightarrow1.87683\times10^{6}\text{m}/\text{s}\\
0. & \text{elsewhere}
\end{cases},\label{eq:dis_order}
\end{eqnarray}
where the phase grid volume $dV=dxdydzdp_{x}dp_{y}dp_{z}$, and $|\mathbf{p}_{i}|\leftrightarrow1.87683\times10^{6}\text{m}/\text{s}$
means the momentum grid $|\mathbf{p}_{j}|$ is obtained via the velocity
$1.87683\times10^{6}\text{m}/\text{s}$. $i\in\{1,...,M\}$ and $,j\in\{1,...,N\}$
are the indices of the discretized spatial and momentum grids (suppose
we have divided the entire phase space into $M\times N$ grids). To
see the pure effects of the diffusion in drift term, we have neglected
all the EM fields in the simulation. The same initial condition is
also simulated by the package JefiPIC\citep{Junjie2022b}, which is
based on the particle simulation method.

\begin{figure}
\begin{centering}
\includegraphics{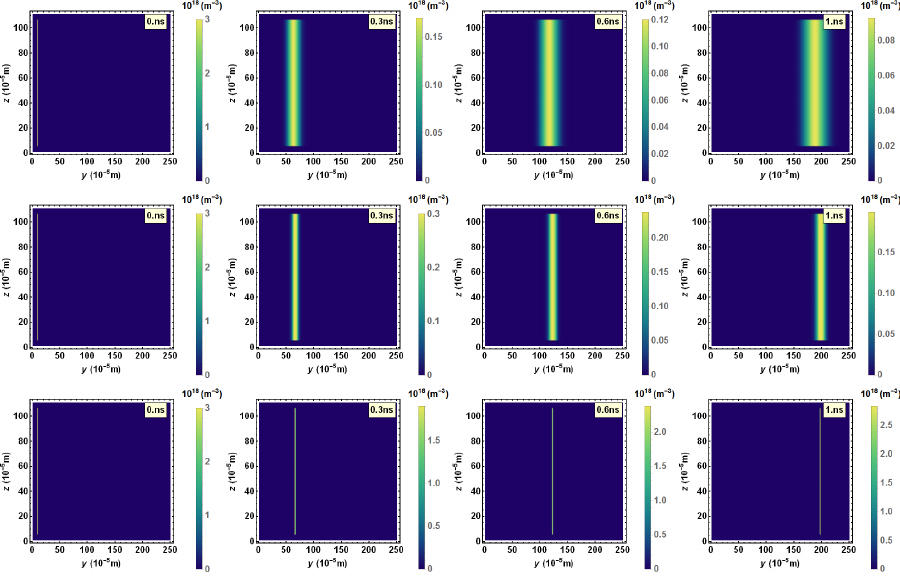}
\par\end{centering}
\caption{Effects of diffusion. All EM fields are set to be zero. The three
lines correspond to the results from RBG-Maxwell 1st order, RBG-Maxwell
2ed order, and JefiPIC.\label{fig:Effects-of-diffusions.}}

\end{figure}

In Fig. \ref{fig:Effects-of-diffusions.}, we present the acquired
particle distributions in the $yoz$ plane. As no electromagnetic
fields are present, the particles are anticipated to move along the
$y$-axis. Due to the numerical diffusion inherent in the finite difference
method, we observe significant diffusion in the first order scheme.
The second order scheme exhibits less diffusion and is hence recommended
for relativistic scenarios. Concurrently, the particle simulation
method exhibits minimal diffusion effects, a reasonable outcome given
its fine resolution in spatial coordinates. Generally, all three models
yield acceptable simulation results. However, diffusion effects are
expected to be suppressed when the electromagnetic fields are applied
head-on (see detailed comparisons in Ref. \citep{Junjie2022b}). The
evaluation time for the first and second order schemes for 10,000
steps is approximately 162 and 168 seconds respectively, when performed
on a single Tesla A 100 card.

\paragraph{Particle distributions and the electric fields}

\begin{figure}

\begin{centering}
\includegraphics[scale=0.4]{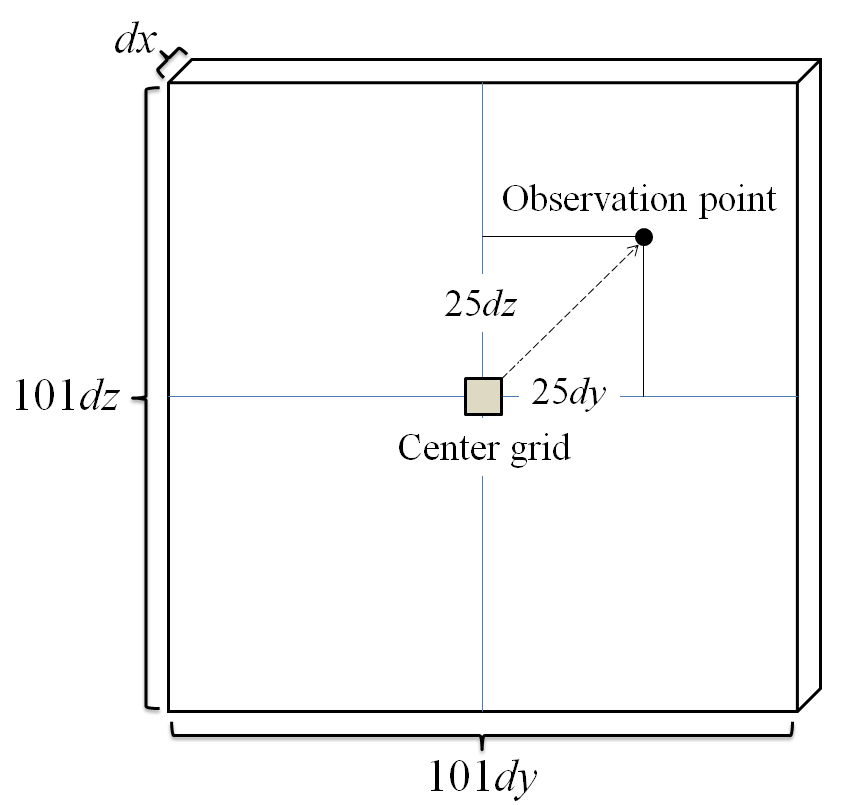}
\par\end{centering}
\caption{Schematic illustration of the initial configuration of the electron
plasma for particle distributions and electric fields. The center
grid is of size $dx\times dy\times dz$. $[0,25,25]$ grids away from
the center grid, an observation point is chosen to record the electric
field along the $y$ direction. \label{fig:Schematic-illustration-of}}

\end{figure}

The electrons are initially positioned at the central grid of the
spatial domain. The 2D domain is segmented into $101\times101$ spatial
grids, each with dimensions of $dx=dy=dz=10^{-5}\text{m}$. An observation
point, located at $[0\text{-th},25\text{-th},25\text{-th}]$ grids
away from the center, monitors the time evolution of the electric
field $E_{y}(t)$. This configuration is schematically depicted in
Fig. \ref{fig:Schematic-illustration-of}. The velocities of the electrons
follow a Gaussian distribution
\begin{eqnarray}
f_{e}(p) & = & N_{e}\frac{1}{\sqrt{2\pi\sigma^{2}}}e^{-\frac{(p-p_{\text{ave}})^{2}}{2\sigma^{2}}},\label{eq:electron_dis}
\end{eqnarray}
where the momentum $p=|\mathbf{p}|$, the average initial momentum
$p_{\text{ave}}=m_{e}v_{\text{ave}}/\sqrt{1-(v_{\text{ave}}/c)^{2}}$,
and $v_{\text{ave}}=1.87864\times10^{6}\text{m}/\text{s}$, $\sigma=0.2p_{\text{ave}}$.
The normalization constant $N_{e}$ is chosen such that $\int d\mathbf{p}f_{e}(p)=N$,
where the total initial electron number $N=31250$. For a smooth spatial
distribution, the particles are distributed in the spatial coordinates
following an exponential expression
\begin{eqnarray}
f_{e}(x) & = & M_{e}e^{-x^{2}},\label{eq:fe_x}
\end{eqnarray}
where the variable $x=|\mathbf{x}|$, and the normalization constant
$M_{e}$ is chosen such that $\int d\mathbf{x}f_{e}(x)=1$. Combining
Eqs. (\ref{eq:electron_dis}) and (\ref{eq:fe_x}), we obtain the initial
distribution function of the electrons
\begin{eqnarray}
f_{e}(\mathbf{x},\mathbf{p},t_{0}) & = & f_{e}(p)f_{e}(x),\label{eq:-12}
\end{eqnarray}
where the initial total particle number is
\begin{eqnarray}
\int d\mathbf{x}d\mathbf{p}f_{e}(\mathbf{x},\mathbf{p},t_{0}) & = & \int d\mathbf{x}d\mathbf{p}f_{e}(p)f_{e}(x)\nonumber \\
 & = & N.\label{eq:total_ini}
\end{eqnarray}

\begin{figure}

\begin{centering}
\includegraphics{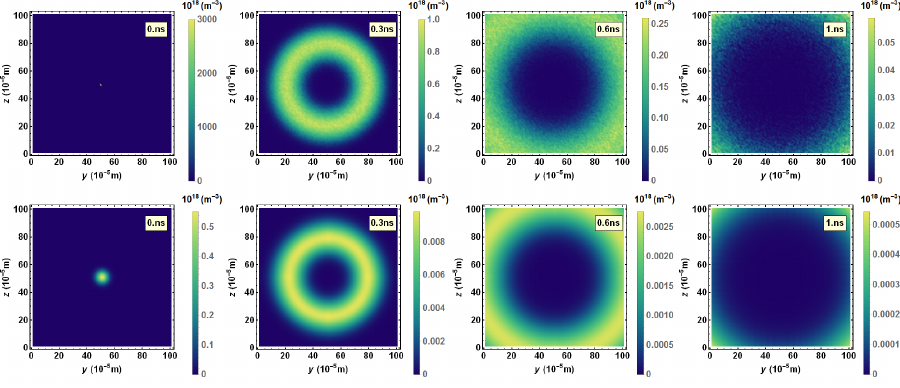}\caption{Snapshots of the particles densities in the $yoz$ plane.\label{fig:Snapshots-of-the}
The first and second rows correspond to the results given by JefiPIC
and RBG-Maxwell. In JefiPIC the particles are initially located in
the center grid as depicted by Fig. \ref{fig:Schematic-illustration-of},
while in RBG-Maxwell we have choosen a particle distribution where
most particles are located in the center grid. The two senarios are
in general equivalent if we neglect the diffusions at the first few
time steps.}
\par\end{centering}
\end{figure}

\begin{figure}

\begin{centering}
\includegraphics[scale=0.3]{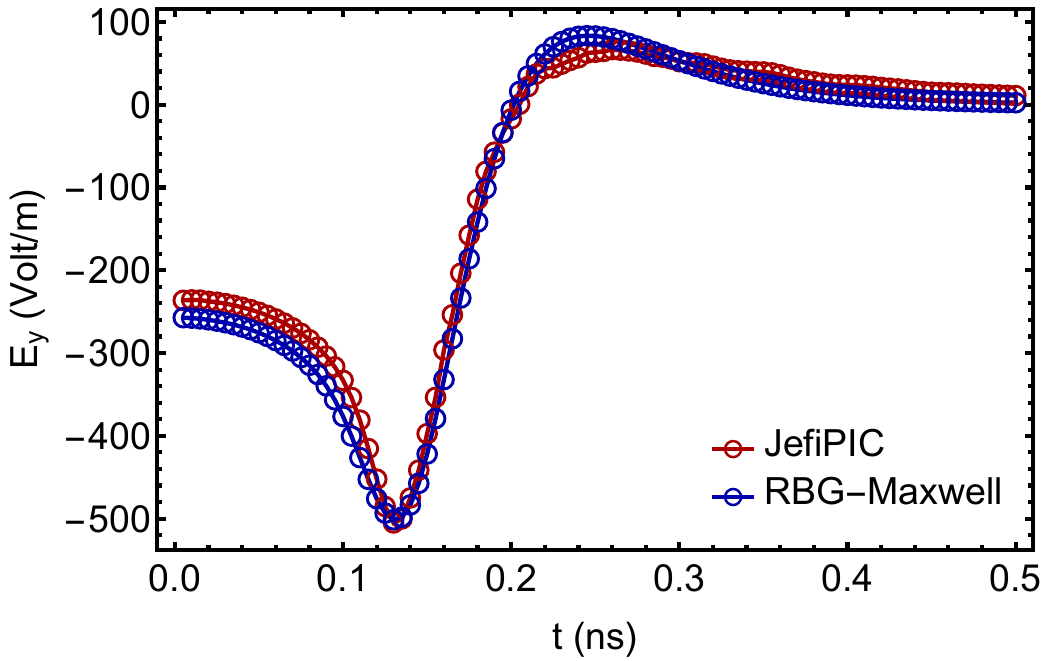}\includegraphics[scale=0.3]{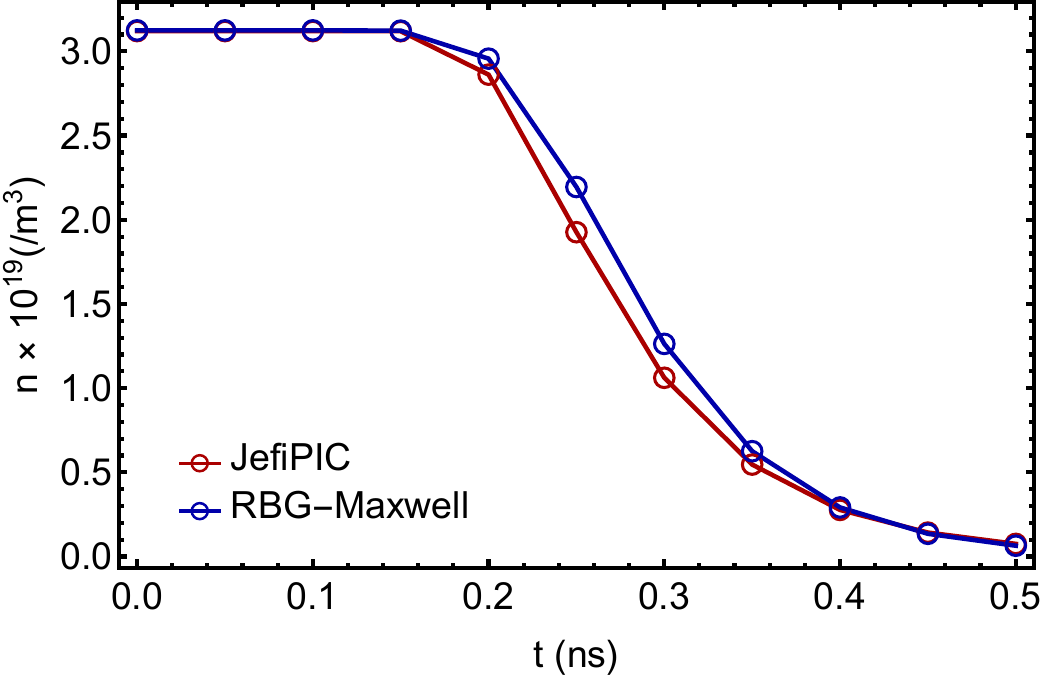}\includegraphics[scale=0.3]{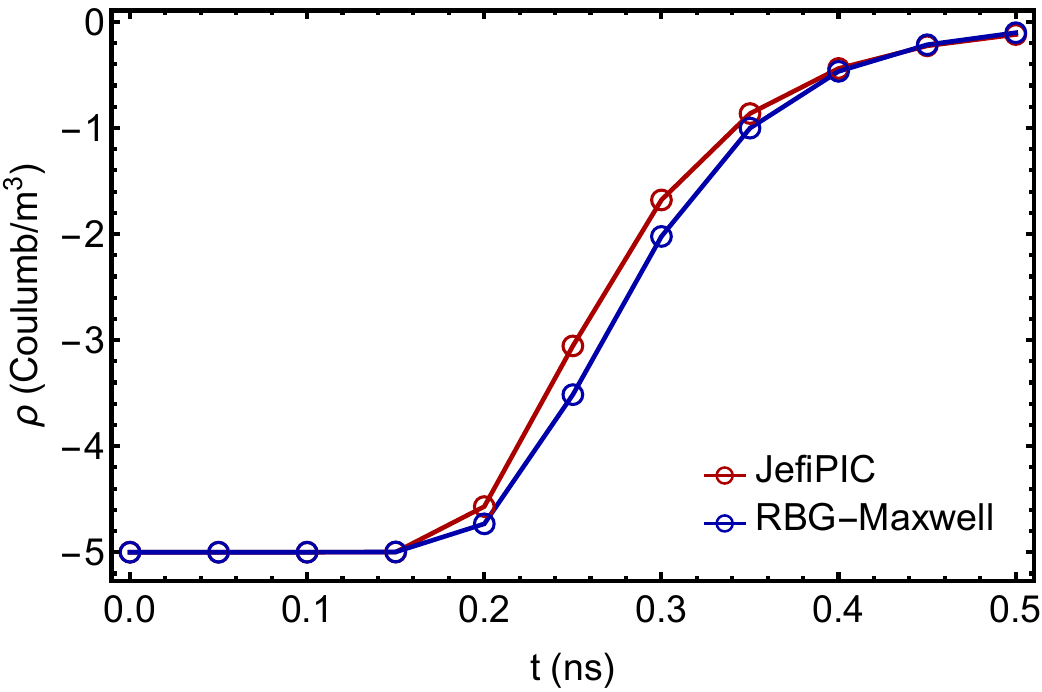}\caption{Comparisons of electric field $E_{y}$, total number density $n$
and total charge density $\rho$.\label{fig:Comparisons-of-electric}
$E_{y}$ is the measured electric field in the $y$ direction at the
observation point. $n$ denote the total number of the particle densities,
i.e., $n=\sum n_{i}$ where $n_{i}$ is the number density in the
$ith$ grid. $\rho$ is the total number of charge densities and $\rho=Qn$,
where $Q$ is the electric charge of an electron.}
\par\end{centering}
\end{figure}

Fig. \ref{fig:Snapshots-of-the} depicts the snapshots of the particle
densities in the $yoz$ plane. A similar model is also executed using
JefiGPU (a particle simulator) to cross-validate our results. It is
evident that the particle densities obtained by both codes are generally
in alignment. To further compare the outcomes of the two packages,
we also present the measured electric field at the observation point,
the total number densities, and total charge densities in Fig. \ref{fig:Comparisons-of-electric}.
The results indicate a perfect correlation between the two packages.

\subsubsection{Test of the collision term}

We set the parameters $\hbar=c=\epsilon_{0}=1$ and $\lambda=1.6\times10^{28}$
in NU. Two particle species $a$ and $b$ with the same mass 0.3 are
chosen. For a box calculation, we restrict the particles in a spatial
box of size $\text{x}_{\text{range}}\times\text{\ensuremath{\text{y}_{\text{range}}}}\times\text{\ensuremath{\text{z}_{\text{range}}}}=[\frac{-6}{0.197},\frac{6}{0.197}]\times[\frac{-6}{0.197},\frac{6}{0.197}]\times[\frac{-6}{0.197},\frac{6}{0.197}]$
with grid size $n_{x}\times n_{y}\times n_{z}=1\times1\times1$. The
distribution functions $f_{a}$ and $f_{b}$ are confined in a six
dimensional phase space of grid size $n_{x}\times n_{y}\times n_{z}\times n_{p_{x}}\times n_{p_{y}}\times n_{p_{z}}=1\times1\times1\times35\times35\times35$.
The momentum ranges are $p_{x,\text{range}}\times p_{y,\text{range}}\times p_{z,\text{range}}=[-2,2]\times[-2,2]\times[-2,2]$.
Here, species $a$ and $b$ are both classical particles, so we expect
that the particles will obey the Boltzmann distribution at thermal
equilibrium (for readers interested in the thermalization of Bosons
and Fermions in BRG-Maxwell, please refer to Ref. \citep{Zhang2020}). 

To see this, we choose the initial distribution function that is far
from thermal equilibrium
\begin{eqnarray}
f & = & f_{0}\theta(|\mathbf{p}|-Q_{s}),\label{eq:step_func}
\end{eqnarray}
where the $\theta$ function is defined as
\begin{eqnarray}
\theta(|\mathbf{p}|-Q_{s}) & = & \begin{cases}
f_{0} & \text{if}\ |\mathbf{p}|>Q_{s}\\
0 & \text{else}
\end{cases}.\label{eq:-11}
\end{eqnarray}
We set $Q_{s}=1$, $f_{0,a}=0.2$ and $f_{0,b}=0.4$ in the calculation.

There are three types of collisions considered, i.e., $a+a\rightarrow a+a$,
$b+b\rightarrow b+b$ and $a+b\rightarrow a+b$. The corresponding
differential cross sections (see definition in Eq. (\ref{eq:ds_M_2_2}))
take the following values
\begin{eqnarray}
\frac{d\sigma}{d\Omega}|_{a+a\rightarrow a+a} & = & 1,\nonumber \\
\frac{d\sigma}{d\Omega}|_{b+b\rightarrow b+b} & = & 20,\nonumber \\
\frac{d\sigma}{d\Omega}|_{a+b\rightarrow a+b} & = & 0.01.\label{eq:dsigma_domega}
\end{eqnarray}
In the evolution, we use time step $dt=0.0002/0.197$ and total number
of time steps $n_{\text{total}}=25000$. The configuration poses a
strong collisional interaction for particle species $b$.

\begin{figure}

\begin{centering}
\includegraphics[scale=0.5]{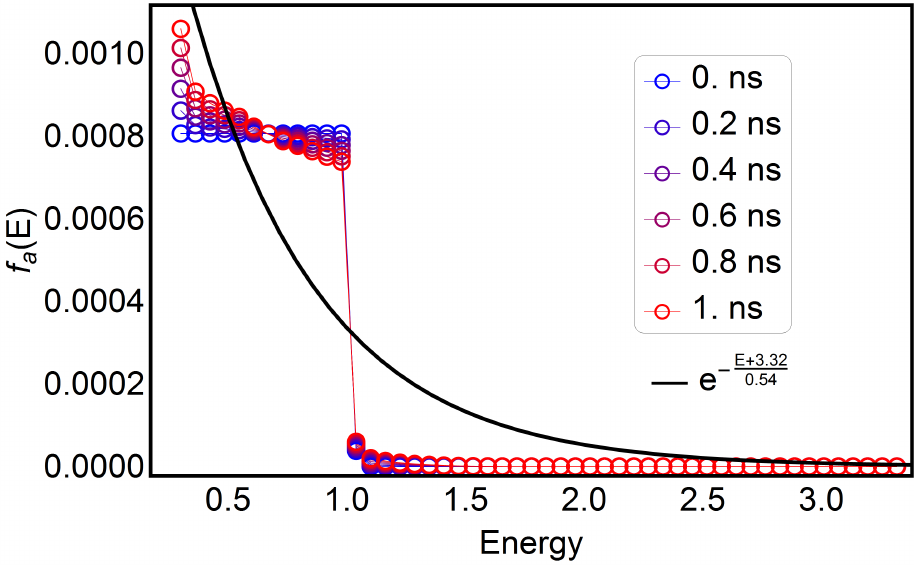}\includegraphics[scale=0.5]{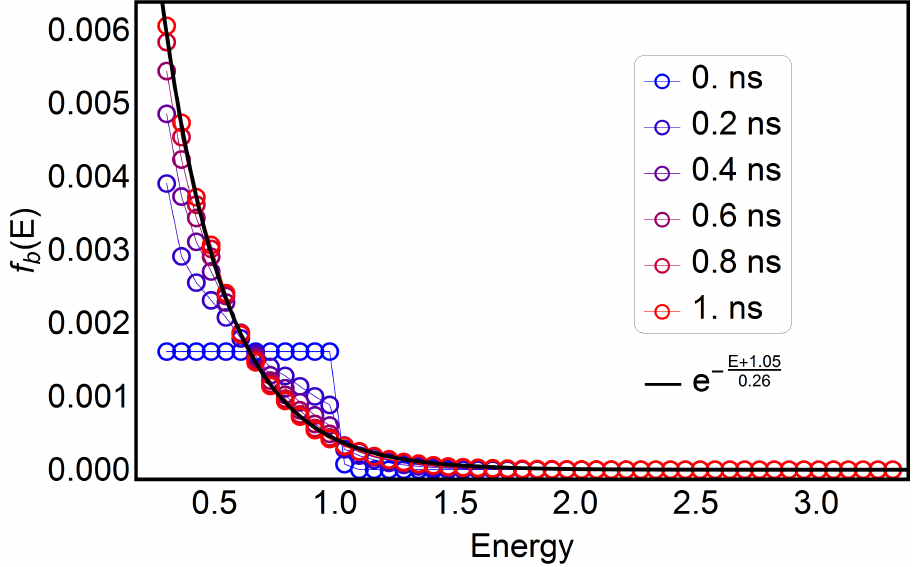}
\par\end{centering}
\caption{Time snapshots of the distribution functions for particle species
$a$ and \textbf{$b$}. The fitted thermal distributions are presented
in the solid lines.\label{fig:Time-snapshots-of}}

\end{figure}

In Fig. \ref{fig:Time-snapshots-of}, we depict the snapshots of the
particle distributions. We can see that the distribution function
of species $b$ gradually achieves the Boltzmann distribution, while
species $a$ is still away from the thermal state. The calculation
takes 0.32 hours on 1 Tesla A 100 card.

\subsubsection{Performance on multi-GPUs}

\begin{figure}
\begin{centering}
\includegraphics[scale=0.4]{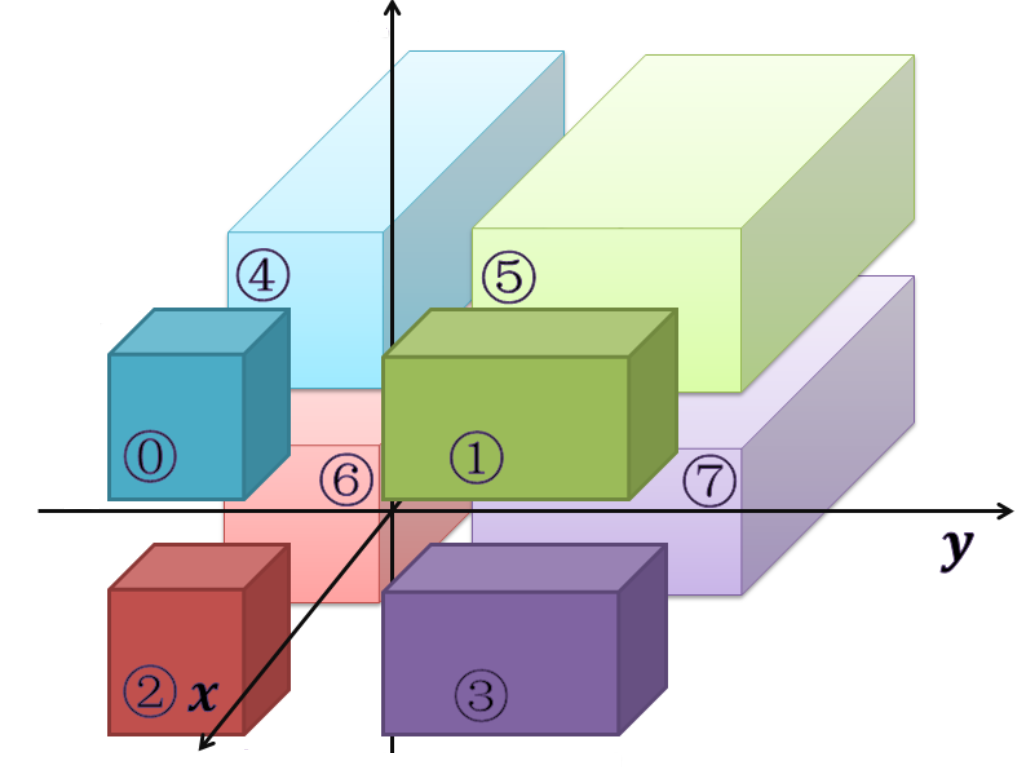}
\par\end{centering}
\caption{Division of the spatial domains. The numbers in the circles denote
the indices of the sub-domains. \label{fig:Division-of-the}}

\end{figure}

We assess the performance of the RBG-Maxwell on multi-GPUs in a more
practical scenario, specifically, the quark-gluon plasma, which is
the original context of the framework. In the simulation of quark-gluon
matter, there are seven particle species interacting with each other,
namely: u, d, and s quarks, their anti-quarks, and gluons. The potential
collision types and the corresponding matrix elements can be found
in Appendix \ref{subsec:Matrix-elements-used}. This intricate collisional
quark-gluon plasma system which can be seamlessly extended to other plasma
systems.
Here, we use 8 NVIDIA A 100 cards in total. The spatial domain is
divided into 8 sub-domains (Fig. \ref{fig:Division-of-the}). Adjacent
sub-domains share two layers as the exchanging boundaries. For convenience,
the number of grid sizes for all sub-domains are set to be the same.
In Tab. \ref{tab:Performance-of-the} we conclude the memory consumption
and the evaluation time for 10 time steps along with different grid
numbers. From Tab. \ref{tab:Performance-of-the} it can be seen that the maximum number of total phase space grids on 8 GPU cards is about
$\sim31^{3}\times11^{3} \approx 4\times10^{7}$. In real practice, we recommend
the use of grid size $n_{x},n_{y},n_{z},n_{p_{x}},n_{p_{y}},n_{p_{z}}=21,21,21,11,11,11$.
Under this configuration, the total evaluation time of 10000 steps
is about a week, which is acceptable in most heavy load computations.
\begin{table}
\begin{centering}
\caption{Performance of the RBG-Maxwell framework on 8 GPU cards. The memory
occupancy is an average of all cards, and the evaluation time is measured
for 10 time steps. Each GPU card has the maximum of 40 GB memory.
\label{tab:Performance-of-the}}
\par\end{centering}
\begin{centering}
\begin{tabular}{cccc}
\hline 
$n_{x},n_{y},n_{z}$ & $n_{p_{x}},n_{p_{y}},n_{p_{z}}$ & Memory occupancy (MB) & Evlaution time (hour)\tabularnewline
\hline 
21,21,21 & 11,11,11 & 18932 & 0.2\tabularnewline
\hline 
21,21,21 & 15,15,15 & 21313 & 0.43\tabularnewline
\hline 
31,31,31 & 11,11,11 & 26766 & 0.65\tabularnewline
\hline 
21,21,21 & 21,21,21 & Out of memory & -\tabularnewline
\hline 
\end{tabular}
\par\end{centering}
\end{table}

\vspace{1.5em}

\section{Conclusion}

In this study, we have detailed the RBG-Maxwell framework, a first-principle
based relativistic collisional plasma simulator, designed for large-scale
GPU clusters. We have outlined the essential equations, numerical
algorithms, implementation details, and key testing outcomes of the
framework. For those interested in utilizing this framework for practical
problem-solving, please visit our introductory webpage:
{[}\textcolor{magenta}{https://Juenjie.github.io or https://sunminmgyan.github.io}{]}.

However, our work is not without its limitations. The current version
of the RBG-Maxwell framework is resource-intensive, typically requiring
eight GPU cards for practical applications. Despite our use of the
Natural Unit to convert all physical quantities into the numerical
range of GPU precision, certain plasma systems still necessitate careful
calibration of the physical quantities. For instance, near space plasma
conditions can present significant variations in velocities, masses,
charges, spatial coordinates, and cross-sections. Additionally, our
current framework struggles to cope with plasma systems that involve
complex interactions with liquid and solid materials (such as the
creation of an electromagnetic pulse, or the interaction of electromagnetic
fields with the walls of the microwave tube). We intend to overcome
these challenges in our future work, and plan to integrate cutting-edge
deep learning techniques to boost the overall performance of the framework.

\section{Acknowledgments}

We extend our gratitude to Xin-Li Sheng from Central China Normal
University for valuable discussions on the formalism, and to Shi Pu
from the University of Science and Technology of China for insights
on the physical implications of our work. This research has been funded
by the National Science Foundation of China under grant number 12105227.

\section*{Appendix}

\subsection*{Integration involving the Dirac delta\label{subsec:Introduction-of-the}}

Dirac delta function is a generalized function which can be loosely
thought of as a function on the real line which is zero everywhere
except at the origin, where it is infinite,
\begin{eqnarray}
\delta(x) & = & \left\{ \begin{array}{c}
+\infty,\ x=0\\
0,\ \ \ \ \ \ x\neq0
\end{array}\right..\label{eq:defdelta}
\end{eqnarray}
Delta function has the translation property
\begin{eqnarray}
\int_{-\infty}^{\infty}f(x)\delta(x-X)dx & = & f(X),\label{eq:relationdelta}
\end{eqnarray}
from which we can work out the delta function corresponding to the
momentum integration,
\begin{eqnarray}
 &  & \int d^{3}\mathbf{k}_{2}\delta^{(3)}(\mathbf{k}_{1}+\mathbf{k}_{2}-\mathbf{k}_{3}-\mathbf{p})g(\mathbf{k}_{2})\nonumber \\
 & = & g(\mathbf{k}_{3}+\mathbf{p}-\mathbf{k}_{1})\vert_{\mathbf{k}_{1}+\mathbf{k}_{2}-\mathbf{k}_{3}-\mathbf{p}=0}\text{,}\label{eq:property1}
\end{eqnarray}
where $g(\mathbf{k}_{2})$ represents an arbitrary function with variable
$\mathbf{k}_{2}$.

For composite delta function $\delta(g(x))$, if $g$ has a real root
$x_{0}$, i.e., $g(x_{0})=0$, then
\begin{eqnarray}
\delta(g(x)) & = & \frac{\delta(x-x_{0})}{|g^{\prime}(x_{0})|},\label{eq:-1}
\end{eqnarray}
where we require the denominator $|g^{\prime}(x_{0})|\neq0$. For
continuously differentiable function $g(x)$, if it has more than
one roots, the composition $\delta(g(x))$ is
\begin{eqnarray}
\delta(g(x)) & = & \sum_{i}\frac{\delta(x-x_{i})}{|g^{\prime}(x_{i})|},\label{eq:-2}
\end{eqnarray}
where the sum extends over all different roots. 

Now we can work out the momentum integration involving $\delta^{(4)}(k_{1}+k_{2}-k_{3}-p)$.
From Eq. (\ref{eq:property1}) we obtain
\begin{eqnarray}
 &  & \int\prod_{i=1}^{3}d^{3}\mathbf{k}_{i}\delta^{(4)}(k_{1}+k_{2}-k_{3}-p)\nonumber \\
 & = & \int\prod_{i=1}^{3}d^{3}\mathbf{k}_{i}\delta^{(3)}(\mathbf{k}_{1}+\mathbf{k}_{2}-\mathbf{k}_{3}-\mathbf{p})\\
 &  & \times\delta(k_{1}^{0}+k_{2}^{0}-k_{3}^{0}-p^{0})\nonumber \\
 & = & \int d^{3}\mathbf{k}_{1}d^{3}\mathbf{k}_{3}\nonumber \\
 &  & \times\delta(k_{1}^{0}+k_{2}^{0}-k_{3}^{0}-p^{0})\vert_{\mathbf{k}_{2}=\mathbf{k}_{3}+\mathbf{p}-\mathbf{k}_{1}}.\label{eq:-3}
\end{eqnarray}
Since the function 
\begin{eqnarray}
g(k_{1}^{z}) & \equiv & k_{1}^{0}+k_{2}^{0}|_{\mathbf{k}_{2}=\mathbf{k}_{3}+\mathbf{p}-\mathbf{k}_{1}}-k_{3}^{0}-p^{0}\nonumber \\
 & = & \sqrt{(k_{1}^{x})^{2}+(k_{1}^{y})^{2}+(k_{1}^{z})^{2}+m^{2}c^{2}}\nonumber \\
 &  & +k_{2}^{0}|_{\mathbf{k}_{2}=\mathbf{k}_{3}+\mathbf{p}-\mathbf{k}_{1}}-k_{3}^{0}-p^{0}\label{eq:-4}
\end{eqnarray}
has two roots 
\begin{eqnarray}
k_{1z}^{\pm} & = & \mathrm{Root}[k_{1}^{0}+k_{2}^{0}|_{\mathbf{k}_{2}=\mathbf{k}_{3}+\mathbf{p}-\mathbf{k}_{1}}-k_{3}^{0}-p^{0}=0],\label{eq:-5}
\end{eqnarray}
from the composition rule in Eq. (\ref{eq:-2}), we have
\begin{eqnarray}
 &  & \delta(k_{1}^{0}+k_{2}^{0}|_{\mathbf{k}_{2}=\mathbf{k}_{3}+\mathbf{p}-\mathbf{k}_{1}}-k_{3}^{0}-p^{0})\nonumber \\
 & = & \sum_{i=\pm}\frac{1}{|J(k_{1z}^{i})|}\delta(k_{1z}-k_{1z}^{i}),\label{eq:delta replace-1-1}
\end{eqnarray}
where the Jacobian function 
\begin{eqnarray}
J(k_{1z}^{\pm}) & = & \frac{\partial g(k_{1}^{z})}{\partial k_{1}^{z}}\vert_{k_{1}^{z}=k_{1z}^{\pm}}\nonumber \\
 & = & \frac{k_{1z}^{\pm}}{(k_{1}^{0})^{\pm}}-\frac{-k_{1z}^{\pm}+k_{3z}+p_{z}}{(k_{2}^{0})^{\pm}}.
\end{eqnarray}
In the above equation,
\begin{eqnarray}
(k_{1}^{0})^{\pm} & = & \sqrt{(k_{1}^{x})^{2}+(k_{1}^{y})^{2}+(k_{1z}^{\pm})^{2}+m_{1}^{2}c^{2}},\nonumber \\
(k_{2}^{0})^{\pm} & = & \sqrt{(\mathbf{k}_{3}+\mathbf{p}-\mathbf{k}_{1})^{2}+m_{2}^{2}c^{2}}\vert_{k_{1}^{z}=k_{1z}^{\pm}},\label{eq:-6}
\end{eqnarray}
where $k_{1z}^{\pm}\equiv\frac{A\pm\sqrt{H}}{B}$, with $A,B,H$ being
the functions of $k_{1}^{x}$,$k_{1}^{y}$,$k_{3}^{x}$,$k_{3}^{y}$,$k_{3}^{z}$,$p_{x}$,$p_{y}$,$p_{z}$,$m_{1}$,$m_{2}$,$k_{3}^{0}$,
and $p^{0}$ (see subsec. \ref{subsec:Finding-the-expression} for
details). 

Substituting Eq. (\ref{eq:delta replace-1-1}) into Eq. (\ref{eq:-3}),
and integrating out $k_{1z}$ using the translation rule Eq. (\ref{eq:relationdelta}),
we have

\begin{align}
 & \int\prod_{i=1}^{3}d^{3}\mathbf{k}_{i}\delta^{(4)}(k_{1}+k_{2}-k_{3}-p)\nonumber \\
= & \int d^{3}\mathbf{k}_{1}d^{3}\mathbf{k}_{3}\delta(k_{1}^{0}+k_{2}^{0}-k_{3}^{0}-p^{0})\nonumber \\
= & \int d^{3}\mathbf{k}_{3}dk_{1}^{x}dk_{1}^{y}\sum_{i=\pm}\frac{1}{|J(k_{1z}^{i})|},\label{eq:work out delta E-1}
\end{align}
which is the expression of Eq. (\ref{eq:work out delta E}).

\subsection*{Finding the expression of $k_{1z}^{\pm}$\label{subsec:Finding-the-expression}}

The expression of $k_{1z}^{\pm}$ can be found via Eq. (\ref{eq:-5}).
In the implementation, we write all components of the momenta explicitly,
hence Eq. (\ref{eq:-5}) becomes

\begin{eqnarray}
k_{1z}^{\pm} & = & \mathrm{Root}[k_{1}^{0}+k_{2}^{0}|_{\mathbf{k}_{2}=\mathbf{k}_{3}+\mathbf{p}-\mathbf{k}_{1}}-k_{3}^{0}-p^{0}=0],\nonumber \\
 & = & \text{Root}\left[\sqrt{(k_{1}^{x})^{2}+(k_{1}^{y})^{2}+(k_{1}^{z})^{2}+m_{1}^{2}c^{2}}\right.-k_{3}^{0}-p^{0}\nonumber \\
 &  & +\sqrt{(k_{3}^{x}+p_{x}-k_{1}^{x})^{2}+(k_{3}^{y}+p_{y}-k_{1}^{y})^{2}+(k_{3}^{z}+p_{z}-k_{1}^{z})^{2}+m_{2}^{2}c^{2}}\nonumber \\
 & = & \frac{C1\pm\sqrt{H}}{C2},\label{eq:-7}
\end{eqnarray}
where
\begin{eqnarray*}
C1 & = & -(k_{3}^{z}+p_{z})\\
 &  & \times\left[c^{2}(-m_{1}^{2}+m_{2}^{2})-(k_{3}^{0}+p^{0})^{2}-2k_{1}^{x}(k_{3}^{x}+p_{x})+(k_{3}^{x}+p_{x})^{2}\right.\\
 &  & \left.-2k_{1}^{y}(k_{3}^{y}+p_{y})+(k_{3}^{y}+p_{y})^{2}+(k_{3}^{z}+p_{z})^{2}\right]\\
C2 & = & 2(k_{3}^{0}-\text{\ensuremath{k_{3}^{z}}}+\text{\ensuremath{p^{0}}}-p_{z})(k_{3}^{0}+k_{3}^{z}+\text{\ensuremath{p^{0}}}+p_{z})\\
H & = & (k_{3}^{0}+p^{0})^{2}\\
 &  & \times\left[(k_{3}^{0})^{4}-2(k_{3}^{0})^{2}(k_{3}^{x})^{2}+k3x^{4}-2(k_{3}^{0})^{2}(k_{3}^{y})^{2}+2(k_{3}^{x})^{2}(k_{3}^{y})^{2}+(k_{3}^{y})^{4}-2(k_{3}^{0})^{2}(k_{3}^{z})^{2}\right.\\
 &  & +2(k_{3}^{x})^{2}(k_{3}^{z})^{2}+2(k_{3}^{y})^{2}(k_{3}^{z})^{2}+(k_{3}^{z})^{4}-2c^{2}(k_{3}^{0})^{2}m_{1}^{2}-2c^{2}(k_{3}^{x})^{2}m_{1}^{2}-2c^{2}(k_{3}^{y})^{2}m_{1}^{2}+2c^{2}(k_{3}^{z})^{2}m_{1}^{2}\\
 &  & +c^{4}m_{1}^{4}-2c^{2}(k_{3}^{0})^{2}m_{2}^{2}+2c^{2}(k_{3}^{x})^{2}m_{2}^{2}+2c^{2}(k_{3}^{y})^{2}m_{2}^{2}+2c^{2}(k_{3}^{z})^{2}m_{2}^{2}-2c^{4}m_{1}^{2}m_{2}^{2}+c^{4}m_{2}^{4}\\
 &  & +4(k_{3}^{0})^{3}p^{0}-4k_{3}^{0}(k_{3}^{x})^{2}p^{0}-4k_{3}^{0}(k_{3}^{y})^{2}p^{0}-4k_{3}^{0}(k_{3}^{z})^{2}p^{0}-4c^{2}k_{3}^{0}m_{1}^{2}p^{0}-4c^{2}k_{3}^{0}m_{2}^{2}p^{0}\\
 &  & +6(k_{3}^{0})^{2}(p^{0})^{2}-2(k_{3}^{x})^{2}(p^{0})^{2}-2(k_{3}^{y})^{2}(p^{0})^{2}-2(k_{3}^{z})^{2}(p^{0})^{2}-2c^{2}m_{1}^{2}(p^{0})^{2}-2c^{2}m2^{2}(p^{0})^{2}+4k_{3}^{0}(p^{0})^{3}+(p^{0})^{4}\\
 &  & -4(k_{3}^{0})^{2}k_{3}^{x}p_{x}+4k3x^{3}p_{x}+4k_{3}^{x}(k_{3}^{y})^{2}p_{x}+4k_{3}^{x}(k_{3}^{z})^{2}p_{x}-4c^{2}k_{3}^{x}m_{1}^{2}p_{x}+4c^{2}k_{3}^{x}m_{2}^{2}p_{x}\\
 &  & -8k_{3}^{0}k_{3}^{x}p^{0}p_{x}-4k_{3}^{x}(p^{0})^{2}p_{x}-2(k_{3}^{0})^{2}p_{x}^{2}+6(k_{3}^{x})^{2}p_{x}^{2}+2(k_{3}^{y})^{2}p_{x}^{2}+2(k_{3}^{z})^{2}p_{x}^{2}\\
 &  & -2c^{2}m_{1}^{2}p_{x}^{2}+2c^{2}m_{2}^{2}p_{x}^{2}-4k_{3}^{0}p^{0}p_{x}^{2}-2(p^{0})^{2}p_{x}^{2}+4k_{3}^{x}p_{x}^{3}+p_{x}^{4}-4(k_{3}^{0})^{2}k_{3}^{y}p_{y}\\
 &  & +4(k_{3}^{x})^{2}k_{3}^{y}p_{y}+4(k_{3}^{y})^{3}p_{y}+4k_{3}^{y}(k_{3}^{z})^{2}p_{y}-4c^{2}k_{3}^{y}m_{1}^{2}p_{y}+4c^{2}k_{3}^{y}m_{2}^{2}p_{y}-8k_{3}^{0}k_{3}^{y}p^{0}p_{y}\\
 &  & -4k_{3}^{y}(p^{0})^{2}p_{y}+8k_{3}^{x}k_{3}^{y}p_{x}p_{y}+4k_{3}^{y}p_{x}^{2}p_{y}-2(k_{3}^{0})^{2}p_{y}^{2}+2(k_{3}^{x})^{2}p_{y}^{2}+6k3y^{2}p_{y}^{2}\\
 &  & +2(k_{3}^{z})^{2}p_{y}^{2}-2c^{2}m_{1}^{2}p_{y}^{2}+2c^{2}m_{2}^{2}p_{y}^{2}-4k_{3}^{0}p^{0}p_{y}^{2}-2p0^{2}p_{y}^{2}+4k_{3}^{x}p_{x}p_{y}^{2}+2px^{2}p_{y}^{2}\\
 &  & +4k_{3}^{y}p_{y}^{3}+p_{y}^{4}+4k_{3}^{z}((k_{3}^{z})^{2}+c^{2}(m_{1}^{2}+m_{2}^{2})-(k_{3}^{0}+p^{0})^{2}+(k_{3}^{x}+p_{x})^{2}+(k_{3}^{y}+p_{y})^{2})pz\\
 &  & +2(3(k_{3}^{z})^{2}+c^{2}(m_{1}^{2}+m_{2}^{2})-(k_{3}^{0}+p^{0})^{2}+(k_{3}^{x}+p_{x})^{2}+(k_{3}^{y}+p_{y})^{2})p_{z}^{2}\\
 &  & +4k_{3}^{z}p_{z}^{3}+p_{z}^{4}+4(k_{1}^{x})^{2}(-(k_{3}^{0}+p^{0})^{2}+(k_{3}^{x}+p_{x})^{2}+(k_{3}^{z}+p_{z})^{2})\\
 &  & +4(k_{1}^{y})^{2}(-(k_{3}^{0}+p^{0})^{2}+(k_{3}^{y}+p_{y})^{2}+(k_{3}^{z}+p_{z})^{2})-4k_{1}^{y}(k_{3}^{y}+p_{y})\\
 &  & \times(c^{2}(-m_{1}^{2}+m_{2}^{2})-(k_{3}^{0}+p^{0})^{2}+(k_{3}^{x}+p_{x})^{2}+(k_{3}^{y}+p_{y})^{2}+(k_{3}^{z}+p_{z})^{2})-4k_{1}^{x}(k_{3}^{x}+p_{x})\\
 &  & \left.\times(c^{2}(-m_{1}^{2}+m_{2}^{2})-(k_{3}^{0}+p^{0})^{2}+(k_{3}^{x}+p_{x})^{2}-2k_{1}^{y}(k_{3}^{y}+p_{y})+(k_{3}^{y}+p_{y})^{2}+(k_{3}^{z}+p_{z})^{2})\right].
\end{eqnarray*}

Therefore, $k_{1z}^{\pm}$ are functions of $k_{1}^{x}$,$k_{1}^{y}$,$k_{3}^{x}$,$k_{3}^{y}$,$k_{3}^{z}$,$p_{x}$,$p_{y}$,$p_{z}$,$m_{1}$,$m_{2}$,$k_{3}^{0}$,
and $p^{0}$. 

\subsection*{Relation between cross section and matrix element\label{subsec:Relation-between-cross}}

For $2\leftrightarrow2$ process, we perform the integration of the
differential probability per unit time,
\begin{eqnarray}
 &  & \int d\omega\nonumber \\
 & = & \int c(2\pi\hbar)^{4}\delta^{(4)}(\sum_{i}p_{i}-\sum_{j}p_{j})|M|^{2}V\left[\prod_{a}\frac{\hbar c}{2E_{a}V}\right]\left[\prod_{b}\frac{d^{3}\mathbf{k}_{b}}{(2\pi\hbar)^{3}}\frac{\hbar c}{2E_{b}}\right]\nonumber \\
 & = & \text{\ensuremath{\int}}\delta^{(4)}(\sum_{i}p_{i}-\sum_{j}p_{j})\frac{\hbar^{2}c|M|^{2}}{V64\pi^{2}k_{1}^{0}k_{2}^{0}k_{3}^{0}p^{0}}d^{3}\mathbf{k}_{3}d^{3}\mathbf{p}.\label{eq:}
\end{eqnarray}
Then we integrate out $d^{3}\mathbf{k}_{3}$ using the delta function
$\delta^{(3)}(\mathbf{p}_{i}-\mathbf{p}_{j})$
\begin{eqnarray}
 &  & \int d\omega\nonumber \\
 & = & \text{\ensuremath{\int}}\delta(k_{1}^{0}+k_{2}^{0}-k_{3}^{0}-\sqrt{|\mathbf{p}|^{2}+m_{p}^{2}c^{2}})\frac{\hbar^{2}c|M|^{2}}{V64\pi^{2}k_{1}^{0}k_{2}^{0}k_{3}^{0}p^{0}}|\mathbf{p}|^{2}d|\mathbf{p}|d\Omega.\label{eq:-9}
\end{eqnarray}
Using the identity

\begin{eqnarray*}
\delta(k_{1}^{0}+k_{2}^{0}-k_{3}^{0}-\sqrt{|\mathbf{p}|^{2}+m_{p}^{2}c^{2}}) & = & \frac{p^{0}}{|\mathbf{p}|}\delta(|\mathbf{p}|-\sqrt{(k_{1}^{0}+k_{2}^{0}-k_{3}^{0})^{2}-m_{p}^{2}c^{2}})
\end{eqnarray*}
we have 
\begin{eqnarray}
 &  & \int d\omega\nonumber \\
 & = & \text{\ensuremath{\int}}\frac{p^{0}}{|\mathbf{p}|}\delta(|\mathbf{p}|-\sqrt{(k_{1}^{0}+k_{2}^{0}-k_{3}^{0})^{2}-m_{p}^{2}c^{2}})\frac{\hbar^{2}c|M|^{2}}{V64\pi^{2}k_{1}^{0}k_{2}^{0}k_{3}^{0}p^{0}}|\mathbf{p}|^{2}d|\mathbf{p}|d\Omega\nonumber \\
 & = & \text{\ensuremath{\int}}\frac{\hbar^{2}c|M|^{2}}{V64\pi^{2}k_{1}^{0}k_{2}^{0}k_{3}^{0}p^{0}}|\mathbf{p}|p^{0}d\Omega\nonumber \\
 & \equiv & \int\frac{cI}{k_{1}^{0}k_{2}^{0}V}d\sigma.\label{eq:-8}
\end{eqnarray}
Therefore, we can obtian the relation between the cross section $d\sigma/d\Omega$
and matrix element $M$ comparing the last two lines of Eq. (\ref{eq:-8}),

\begin{eqnarray}
\frac{d\sigma}{d\Omega} & = & \frac{\hbar^{2}|M|^{2}|\mathbf{p}|}{64\pi^{2}k_{3}^{0}I}.\label{eq:-10}
\end{eqnarray}

\subsection*{Matrix elements used in the model calculation\label{subsec:Matrix-elements-used}}

\begin{table}[H]
\begin{centering}
\caption{\label{tab:Matrix-elements-squared} Matrix elements squared for all
$2\rightarrow2$ parton scattering processes in QCD. The helicities
and colors of all initial and final state particles are summed over.
$q_{1}$ ($\bar{q}_{1}$) and $q_{2}$ ($\bar{q}_{2}$) represent
quarks (antiquarks) of different flavors, and $g$ represents the
gluon. $d_{F}$ and $d_{A}$ denote the dimensions of the fundamental
and adjoint representations of $SU_{c}(N)$ gauge group while $C_{F}$
and $C_{A}$ are the corresponding quadratic Casimirs. In a $SU_{c}(3)$
theory with fundamental representation fermions, $d_{F}=C_{A}=3$,
$C_{F}=4/3$, and $d_{A}=8$. The infrared divergence is suppressed
by introducing a regulator in the denominator\citep{Arnold2003a,Chen2013,Zhang1998a}.}
\par\end{centering}
\centering{}%
\begin{tabular}{c|c}
\hline 
$ab\rightarrow cd$ & $\left\vert M_{a\left(k_{1}\right)b\left(k_{2}\right)\rightarrow c\left(k_{3}\right)d\left(p\right)}\right\vert ^{2}$\tabularnewline
\hline 
$\begin{array}{c}
q_{1}q_{2}\rightarrow q_{1}q_{2}\\
\bar{q}_{1}q_{2}\rightarrow\bar{q}_{1}q_{2}\\
q_{1}\bar{q}_{2}\rightarrow q_{1}\bar{q}_{2}\\
\bar{q}_{1}\bar{q}_{2}\rightarrow\bar{q}_{1}\bar{q}_{2}
\end{array}$ & $8g^{4}\dfrac{d_{F}^{2}C_{F}^{2}}{d_{A}}\left(\dfrac{s^{2}+u^{2}}{(t-m_{g}^{2})^{2}}\right)$\tabularnewline
\hline 
$\begin{array}{c}
q_{1}q_{1}\rightarrow q_{1}q_{1}\\
\bar{q}_{1}\bar{q}_{1}\rightarrow\bar{q}_{1}\bar{q}_{1}\\
\\
\end{array}$ & $8g^{4}\dfrac{d_{F}^{2}C_{F}^{2}}{d_{A}}\left(\dfrac{s^{2}+u^{2}}{(t-m_{g}^{2})^{2}}+\dfrac{s^{2}+t^{2}}{(u-m_{g}^{2})^{2}}\right)$$+16g^{4}d_{F}C_{F}\left(C_{F}-\frac{C_{A}}{2}\right)\dfrac{s^{2}}{(t-m_{g}^{2})(u-m_{g}^{2})}$\tabularnewline
\hline 
$\begin{array}{c}
q_{1}\bar{q}_{1}\rightarrow q_{1}\bar{q}_{1}\\
\\
\end{array}$ & $8g^{4}\dfrac{d_{F}^{2}C_{F}^{2}}{d_{A}}\left(\dfrac{s^{2}+u^{2}}{(t-m_{g}^{2})^{2}}+\dfrac{u^{2}+t^{2}}{s^{2}}\right)$$+16g^{4}d_{F}C_{F}\left(C_{F}-\frac{C_{A}}{2}\right)\dfrac{u^{2}}{(t-m_{g}^{2})s}$\tabularnewline
\hline 
$\begin{array}{c}
q_{1}\bar{q}_{1}\rightarrow q_{2}\bar{q}_{2}\\
\\
\end{array}$ & $8g^{4}\dfrac{d_{F}^{2}C_{F}^{2}}{d_{A}}\dfrac{t^{2}+u^{2}}{s^{2}}$\tabularnewline
\hline 
$\begin{array}{c}
q_{1}\bar{q}_{1}\rightarrow gg\\
\\
\end{array}$ & $8g^{4}d_{F}C_{F}^{2}\left(\dfrac{u}{(t-m_{g}^{2})}+\dfrac{t}{(u-m_{g}^{2})}\right)-8g^{4}d_{F}C_{F}C_{A}\left(\dfrac{t^{2}+u^{2}}{s^{2}}\right)$\tabularnewline
\hline 
$\begin{array}{c}
q_{1}g\rightarrow q_{1}g\\
\bar{q}_{1}g\rightarrow\bar{q}_{1}g
\end{array}$ & $-8g^{4}d_{F}C_{F}^{2}\left(\dfrac{u}{s}+\dfrac{s}{(u-m_{g}^{2})}\right)+8g^{4}d_{F}C_{F}C_{A}\left(\dfrac{s^{2}+u^{2}}{(t-m_{g}^{2})^{2}}\right)$\tabularnewline
\hline 
$\begin{array}{c}
gg\rightarrow gg\\
\\
\end{array}$ & $16g^{4}d_{A}C_{A}^{2}\left(3-\dfrac{su}{(t-m_{g}^{2})^{2}}-\dfrac{st}{(u-m_{g}^{2})^{2}}-\dfrac{tu}{s^{2}}\right)$\tabularnewline
\hline 
\end{tabular}
\end{table}

\vspace{1.5em}

\nocite{*}
\bibliography{draft-20230624}

\end{document}